\documentclass[twocolumn,aps,nofootinbib]{revtex4-1}
\usepackage{graphicx}
\usepackage{epstopdf}
\usepackage{latexsym}
\usepackage{amssymb}
\usepackage{amsmath}
\usepackage{color}
\usepackage{mathrsfs}
\usepackage{xparse}
\usepackage{float}
\usepackage{mathtools}
\usepackage[center]{subfigure}
\usepackage{placeins}
\usepackage{multirow}
\usepackage{tabularx}
\usepackage{float}

\begin{document}
\renewcommand\arraystretch{2}
\newcommand{\bq}{\begin{equation}}
\newcommand{\eq}{\end{equation}}
\newcommand{\bqn}{\begin{eqnarray}}
\newcommand{\eqn}{\end{eqnarray}}
\newcommand{\nb}{\nonumber}
\newcommand{\lb}{\label}
\newcommand{\cb}{\color{blue}}
\newcommand{\cc}{\color{cyan}}
\newcommand{\cm}{\color{magenta}}
\newcommand{\rc}{\rho^{\scriptscriptstyle{\mathrm{I}}}_c}
\newcommand{\rd}{\rho^{\scriptscriptstyle{\mathrm{II}}}_c} 
\NewDocumentCommand{\evalat}{sO{\big}mm}{%
  \IfBooleanTF{#1}
   {\mleft. #3 \mright|_{#4}}
   {#3#2|_{#4}}%
}

\newcommand{\PRL}{Phys. Rev. Lett.}
\newcommand{\PL}{Phys. Lett.}
\newcommand{\PR}{Phys. Rev.}
\newcommand{\CQG}{Class. Quantum Grav.}
\newcommand{\parallelsum}{\mathbin{\!/\mkern-5mu/\!}}
\renewcommand\arraystretch{2}
 \newcommand{\subbq}{\begin{subequations}}
 \newcommand{\subeq}{\end{subequations}}
\newcommand{\rcone}{\rho^{\scriptscriptstyle{\mathrm{I}}}_c}
\newcommand{\La}{\Lambda}
\newcommand{\va}{\scriptscriptstyle}
\newcommand{\be}{\nopagebreak[3]\begin{equation}}
\newcommand{\ee}{\end{equation}}
\newcommand{\sign}{\text{sign}}

\newcommand{\ba}{\nopagebreak[3]\begin{eqnarray}}
\newcommand{\ea}{\end{eqnarray}}

\newcommand{\la}{\label}
\newcommand{\n}{\nonumber}
\newcommand{\su}{\mathfrak{su}}
\newcommand{\SU}{\mathrm{SU}}
\newcommand{\U}{\mathrm{U}}

\def\be{\nopagebreak[3]\begin{equation}}
\def\ee{\end{equation}}
\def\ba{\nopagebreak[3]\begin{eqnarray}}
\def\ea{\end{eqnarray}}
\newcommand{\f}{\frac}
\def\rmd{\rm d}
\def\lp{\ell_{\rm Pl}}
\def\d{{\rm d}}
\def\fe{\mathring{e}^{\,i}_a}
\def\fw{\mathring{\omega}^{\,a}_i}
\def\fq{\mathring{q}_{ab}}
\def\t{\tilde}

\def\db{\delta_b}
\def\dc{\delta_c}
\def\T{\mathcal{T}}
%%THESE ARE NEW MACROS USED IN APPENDIX A
\def\GammaE{\Gamma_{\rm ext}}
\def\GammaEb{\bar\Gamma_{\rm ext}}
\def\GammaEh{\hat\Gamma_{\rm ext}}
\def\Hee{H_{\rm eff}^{\rm ext}}
\def\H{\mathcal{H}}

\newcommand{\R}{\mathbb{R}}
\providecommand{\apjs}{ApJS}

%%%%%%%%%%%%%%%%%%%%%%%%%%%%%%%%%%%%%%%%%%%%%%%%%%%%%%%%%%%%%%%%%%%%%%%%%%%%%%%%%%%%%%%%%%%%%%%%%%%

\title{Effects of the ekpyrotic mechanism on  inflationary phase  in loop quantum cosmologies} 

\author{Christian Brown$^{a}$}
\email{Christian$\_$Brown4@baylor.edu}

\author{Jared Fier$^{b}$}
\email{Jared$\_$Fier@baylor.edu}

\author{Brian Phillips$^{a}$}
\email{Brian$\_$Phillips1@baylor.edu}

\author{Gerald Cleaver$^{a}$}
\email{Gerald$\_$Cleaver@baylor.edu}

\author{Anzhong Wang$^{b}$ \footnote{The corresponding author}}
\email{anzhong$\_$wang@baylor.edu; the corresponding author}

\affiliation{$^{a}$ EUCOS-CASPER,  Department of Physics and Astronomy, Baylor University, Waco, TX 76798-7316, USA\\
$^{b}$ GCAP-CASPER, Department of Physics and Astronomy, Baylor University, Waco, TX 76798-7316, USA\\}

\date{\today}
\begin{abstract}

In bouncing cosmological models - whether classical or quantum - the big bang singularity is replaced by a regular bounce. A well‑known challenge is controlling the growth of shear during the contracting phase, as the shear scales as $a^{-6}$ toward the bounce, where $a$ is the average expansion factor of the universe. A common solution is to introduce a scalar field with an ekpyrotic‑like potential that becomes negative near the bounce, giving an effective equation of state greater than one, thereby dominating the shear and allowing a homogeneous, isotropic universe to emerge after the bounce. In this paper, we investigate how the ekpyrotic mechanism affects the inflationary phase in both loop quantum cosmology (LQC) and a modified loop quantum cosmology model (mLQC‑I) - frameworks in which inflation is otherwise generic. We consider a potential consisting of an inflationary part and an ekpyrotic‑like component. Through numerical studies of various models, we find that the ekpyrotic mechanism can significantly influence the subsequent inflation, even though, for appropriately chosen parameters, it dominates near the bounce (with equation of state $> 1$) and successfully resolves the shear problem. As time increases beyond the bounce, the inflationary potential eventually dominates, producing an inflationary phase that can be long enough to address standard hot big bang cosmology problems.
Nevertheless, our numerical findings suggest that fine-tuning may be required. Because the results are numerical, our conclusions are not definitive, and a more systematic analysis will be necessary to fully assess the generality of the observed behavior.

\end{abstract}

\maketitle

\section{
Introduction
}
\renewcommand{\theequation}{1.\arabic{equation}}
\setcounter{equation}{0}

Since its incarnation in 1980 \cite{1981PhRvD..23..347G}, the inflationary paradigm has achieved great success, resolving many long-standing problems of the standard big bang cosmology, and is consistent with all cosmological and astrophysical observations conducted so far \cite{Planck:2018jri,AtacamaCosmologyTelescope:2025blo}. However, the paradigm has also faced some challenges. 
In particular,  it is well-known that this paradigm is sensitive to the ultraviolet (UV) physics, and its successes are tightly contingent on the understanding of this UV physics \cite{Brandenberger:2012aj,Silverstein:2016ggb,Baumann:2014nda}. Typically, if the inflationary phase lasts somewhat longer than the minimal period required to solve the above mentioned problems, the length scales we observe today can originate from modes that are smaller than the Planck length during inflation. Then, the treatment of the underlying quantum field theory on a classical spacetime background becomes questionable, as now the quantum geometric effects are expected to be large, and the space and time cannot be treated  classically  any longer. This is often referred to as  {\em the trans-Planckian problem} of cosmological fluctuations \cite{Brandenberger:2012aj}.

The second problem  is related to the existence of the big bang singularity \cite{Borde:1993xh,Borde:2001nh}, with which it is not clear how to impose initial conditions. Instead, one often ignores the pre-inflationary dynamics and sets the initial conditions at a sufficiently early time so that all the observational modes are inside the Hubble horizon. In the slow-roll inflation scenario, the spacetime becomes almost de Sitter,  and the Bunch-Davies (BD) vacuum becomes a natural choice \cite{Bunch:1978yq}. 
However, it is still an open question on how such a vacuum state can be realized dynamically in the framework of quantum cosmology (QC), considering the fact that a pre-inflationary phase always exists between the Plank and inflation scales, which are about $10^{12}$ orders of magnitude difference in terms of energy densities.  During this phase, particle creations are inevitable. 

It is clear that all the above issues are closely related to QC, a topic that has been extensively studied in the past decades, and various theories have been proposed. Among them are models constructed from string/M-theory \cite{Green_Schwarz_Witten_2012,Becker:2006dvp} and loop quantum gravity (LQG) \cite{Ashtekar:2004eh,Thiemann_2007,Bojowald_2010,Gambini:2011zz,Rovelli:2014ssa}. In particular, in the last two decades, LQG has been rigorously applied to understand singularity resolution in various cosmological models
(for recent reviews, see Refs. \cite{Ashtekar:2011ni,ElizagaNavascues:2020uyf,Li:2023dwy,Agullo:2023rqq}), and    a coherent picture of Planck scale physics has emerged: {\em the big bang singularity is replaced by a quantum bounce, purely due to quantum geometric effects}. This framework is often referred to as loop quantum cosmology (LQC).
In the last couple of years, to understand some ambiguities of LQC, several modified  loop quantum cosmological (mLQC) models have been proposed \cite{Li:2021mop}, including mLQC-I \cite{Li:2018opr,Li:2018fco,Li:2019ipm}, first proposed in \cite{Yang:2009fp} and later systematically developed in \cite{Assanioussi:2018hee,Assanioussi:2019iye}. It is interesting to note that this model can be also obtained by the so-called top-down approach \cite{Dapor:2017rwv,Dapor:2017gdk,Han:2021cwb}.

In all bouncing cosmological models, either classical \cite{Lehners:2008vx,Battefeld:2014uga,Brandenberger:2016vhg} or quantum \cite{Ashtekar:2011ni,Li:2023dwy,Agullo:2023rqq}, a challenging question is how to solve the shear problem. This is because in a contracting phase, shear always grows like $a^{-6}$, which is faster than all matter fields, except for the stiff fluid (or a massless scalar field) that also grow as in the same rate, where $a$ is the average expansion factor. However, even in the latter it is not clear how the stiff fluid can always win over the shear, so that a homogeneous and isotropic universe will develop after the bounce. Shear in homogeneous and anisotropic Bianchi  universes have been extensively investigated in LQC \cite{Ashtekar:2011ni,Li:2023dwy,Agullo:2023rqq} and   various interesting results have been obtained. In particular, in the Bianchi I universe it was found that the shear is always conserved asymptotically \cite{Chiou:2007sp,Ashtekar:2009vc}. Therefore, to solve the shear problem in LQC, one often borrows  the ekpyrotic mechanism (see for example,  \cite{McNamara:2022dmf,Motaharfar:2023hil} and references therein), first introduced in  colliding branes \cite{Khoury:2001wf,Lehners:2008vx} and later generalized to  other bouncing models, including  matter and ekpyrotic/cyclic bounces \cite{Cai:2012va,Brandenberger:2016vhg,Ijjas:2019pyf,Ijjas:2020dws,Ijjas:2021zyf,Tukhashvili:2023itb,Ijjas:2024oqn,Itzhaki:2025gdv}. The basic idea is to  introduce a scalar field with an ekpyrotic-like potential which becomes negative near the bounce, so the effective equation of state (EoS) of the scalar field will be greater than one, so that the scalar field will grow like $\rho_{\phi} \propto a^{-3(1+w)} \;  (w > 1)$, whereby dominates the shear in the bounce region. As a result, a homogeneous and isotropic universe can be developed after the bounce. 
 
 In this paper, we study how the ekpyrotic mechanism affects the inflationary phase in  both LQC and mLQC-I, because in these frameworks the inflation is generic without such a mechanism \cite{Ashtekar:2011rm,Li:2019ipm}. Then, a natural question is whether the inflation is still generic or not after the ekpyrotic mechanism is taken into account. To answer this question, we consider a scalar field with a total potential given by 
 \bqn
 \lb{eq1.1}
 V(\phi) =  V_{\text{ekp}}(\phi) + V_{\text{inf}}(\phi),
 \eqn
 where $V_{\text{ekp}}(\phi)$ denotes an ekpyrotic type potential, and $V_{\text{inf}}(\phi)$ an inflationary potential.
 Clearly, to have the mechanism work, $V_{\text{ekp}}(\phi)$ needs to dominate the evolution of the universe in the contracting phase near the bounce, while after the bounce the inflationary potential $V_{\text{inf}}(\phi)$ will gradually increase and finally dominate the evolution, whereby an inflationary phase is developed. Therefore, the task now reduces to showing that the above mentioned process indeed occurs for a given set of initial conditions. More importantly, the inflationary phase will last long enough to solve the big bang problems, which motivated the proposal of inflation in the first place \cite{1981PhRvD..23..347G}. 

  It must be noted that  in the  matter and ekpyrotic/cyclic bounces, see, for example,  \cite{Lehners:2008vx,Battefeld:2014uga,Brandenberger:2016vhg,Cai:2012va,Ijjas:2019pyf,Ijjas:2020dws,Ijjas:2021zyf,Tukhashvili:2023itb,Ijjas:2024oqn,Itzhaki:2025gdv} and references therein, an inflationary phase is not required. As a matter of fact, one of the main motivations of these models is precisely to replace the inflationary phase by a regular bounce, as the latter naturally solve the big bang singularity and trans-Planck problems. In addition, a matter-dominated contracting phase also leads to a power spectrum that is scale-invariant \cite{Wands:1998yp}. 
  In LQC and mLQC-I, primordial power spectrum   without inflation  has been also studied  \cite{Bojowald:2004kt,Wilson-Ewing:2012lmx,Wilson-Ewing:2013bla,Wilson-Ewing:2015sfx,Li:2021fmu}. However, recently it was  found that the resultant power spectrum is inconsistent with current observations \cite{Li:2020pww}. 

  Therefore, in this paper we shall focus on LQC/mLQC-I models in which inflationary phase still exist, as we mentioned above that inflation is generic in these models when the ekpyrotic mechanism is not taken into account \cite{Ashtekar:2011rm,Li:2019ipm}. 
  In particular, after numerically studying various cases, we find that, by properly choosing the free parameters involved in the models,  the ekpyrotic-like potential indeed dominates the evolution of the universe in the bounce region, during which the EoS  of the scalar field  is larger than one, so the shear problem is resolved. As time continuously increases after the bounce, the inflationary potential  picks up and becomes dominant, whereby an inflationary phase is finally developed. This phase can last long enough in order to solve the cosmological problems of the big bang cosmology.

The rest of the paper is organized as follows:   In Sec. II we give a brief introduction to LQC and mLQC-I, and provide the corresponding Hamiltonian equations. In Sec. III we solve these equations numerically with the total potential given by Eq.(\ref{eq1.1}) for various choices of the parameters involved in the models in the framwork of LQC, while in Sec. IV, we generalize such investigations to mLQC-I. Although the existence of an inflationary phase
with sufficient e-folds sensitively depend on the choices of the free parameters, we do find regions of the parameter phase 
spacetime with non-zero measure that lead to such desirable inflation. However, our numerical findings suggest that fine-tuning may be required. In Sec. V, 
we summarize our main results and provide some concluding remarks.

%%%%%%%%%%%%%%%%%%%%%%%%%%%%%%%%%%%%%%%%%%%%%%%%%%%%%%%%%%

\section{Effective Dynamical Equations}
\renewcommand{\theequation}{2.\arabic{equation}}
\setcounter{equation}{0}
\lb{SecII}
%%%%%%%%%%%%%%%%%%%%%%%%%%%%%%%
%%%%%%%%%%%%%%%%%%%%%%%%%%%%%%

In this section, we provide a summary of the modified Friedmann dynamics in the frameworks of LQC \cite{Ashtekar:2011ni}and  mLQC-I \cite{Li:2018fco}. 

\subsection{Effective Dynamical Equations in LQC}

In the framework of LQC, the dynamics can be obtained from the effective Hamiltonian given by
\bq
\lb{eq2.1}
\mathcal{H}_{\text{LQC}} = - \frac{3v}{8\pi G \gamma^2 }  \frac{\sin^2 (\lambda b)}{\lambda^2} +\mathcal{H_M},
\eq 
where  $G$ is the Newtonian constant, $v \equiv a^3$,  
and $a$ is the expansion factor of the Universe 
\bq
\lb{eq2.2}
ds^2 = - dt^2 + a^2(t)\left(dx^2 + dy^2 +dz^2\right).
\eq 
The variable $b$ denotes the momentum conjugate of $v$ and satisfies the canonical relation
\bq
\lb{eq2.3}
\{b,v\} = 4 \pi G \gamma,
\eq 
where  $\gamma$ is known as the Barbero-Immirzi parameter whose value is set to $\gamma \approx 0.2375$ using black hole thermodynamics in LQG \cite{Meissner:2004ju}. The parameter $\lambda$ is defined as $\lambda^2 \equiv \Delta = 4 \sqrt{3}\pi\gamma\ell^2_\text{pl}$, where $\Delta$ denotes the minimal area gap of the area operator in LQG \cite{Ashtekar:2004eh,Thiemann_2007,Bojowald_2010,Gambini:2011zz,Rovelli_Vidotto_2014,Ashtekar:2017yom}. The matter 
Hamiltonian $\mathcal{H_M}$ is given by
\bq
\lb{eq2.4}
\mathcal{H_M} = v \rho,  
\eq 
where $\rho$ denotes the energy density of the matter field. Then, the Hamiltonian equation for a given physical quantity $A$ of the system
\bq
\lb{eq2.5}
\dot{A} = \left\{A, \mathcal{H}\right\},   
\eq 
yields
\bqn
\lb{eq2.6}
\dot{b} &=& \left\{b, \mathcal{H}_{\text{LQC}}\right\} =  4 \pi G \gamma \frac{\partial \mathcal{H}_{\text{LQC}}}{\partial v}   \nb\\
&=& 4\pi G\gamma\left[- \frac{3}{8\pi G \gamma^2}  \frac{\sin^2 (\lambda b)}{\lambda^2} + \frac{\partial\mathcal{H_M}}{\partial v} \right],\\
\lb{eq2.7}
\dot{v} &=& \left\{v, \mathcal{H}_{\text{LQC}}\right\} = - 4 \pi G \gamma \frac{\partial \mathcal{H}_{\text{LQC}}}{\partial b}   \nb\\
&=& \frac{3v}{2\lambda\gamma}\sin\left(2\lambda b\right).
\eqn 
On the other hand, from the Hamiltonian constraint $\mathcal{H}_{\text{LQC}} \simeq 0$ we find that 
\bq
\lb{eq2.8}
\rho   =  \frac{3}{8\pi G \gamma^2 }  \frac{\sin^2 (\lambda b)}{\lambda^2}. 
\eq 
Inserting the above expression into Eqs.(\ref{eq2.6}) and (\ref{eq2.7}) we find that 
\bqn
\lb{eq2.9}
\dot{b} &=&  -  4 \pi G \gamma \left(\rho + P\right),  \\
\lb{eq2.10}
H^2 &\equiv& \left(\frac{\dot{v}}{3v}\right)^2 = \frac{8\pi G}{3}\rho\left(1 - \frac{\rho}{\rho_c}\right),  
\eqn 
where the pressure $P$ and critical energy density $\rho_c $ are defined respectively as
\bqn
\lb{eq2.11}
P \equiv -  \frac{\partial \mathcal{H_M}}{\partial v}, \quad
\rho_c \equiv   \frac{3}{8\pi G\lambda^2\gamma^2}.  
\eqn 
From Eq.(\ref{eq2.10}) we can see that $\rho \leq \rho_c$, and when  $\rho = \rho_c$ we have   $H = \dot{a}/a = 0$, at which a quantum bounce happens.

For a scalar field $\phi$ with a potential $V(\phi)$, we have 
\bq
\lb{eq2.12}
\mathcal{H_{\phi}} = \frac{p_{\phi}^2}{2v} + v V(\phi),  
\eq 
where $p_{\phi}$ is the momentum conjugate of $\phi$ and satisfies the canonical relation 
\bq
\lb{eq2.13}
\{\phi,p_{\phi}\} = 1. 
\eq 
Then, the Hamiltonian equation (\ref{eq2.5}) yields
\bqn
\lb{eq2.14}
\dot \phi &=& \{\phi,\mathcal{H}\} = \frac{\partial\mathcal{H_{\phi}}}{\partial p_{\phi}} = \frac{p_{\phi}}{v}, \\
\lb{eq2.15}
\dot{p}_{\phi} &=& \{p_{\phi},\mathcal{H}\} = -\frac{\partial\mathcal{H_{\phi}}}{\partial \phi} = - v V_{,\phi},
\eqn
where $V_{,\phi} \equiv dV(\phi)/d\phi$. From the above equations, we find that
\bqn
\lb{eq2.16}
\ddot\phi + 3H \dot\phi + V_{,\phi}(\phi) = 0,  
\eqn
which is nothing but the Klein-Gordon equation. 

On the other hand, from Eqs.(\ref{eq2.4}), (\ref{eq2.11}) and (\ref{eq2.14}) we find that
\bqn
\lb{eq2.17}
\rho_{\phi} &=&  \frac{p_{\phi}^2}{2v^2} +  V(\phi) = \frac{1}{2}\dot{\phi}^2 + V(\phi),\nb\\
P_{\phi} &=&  \frac{p_{\phi}^2}{2v^2} -  V(\phi) = \frac{1}{2}\dot{\phi}^2 - V(\phi).
\eqn
Then, the equation of state (EoS) for the scalar field is given by
\bqn
\lb{eq2.18}
w_{\phi} \equiv \frac{P_{\phi}}{\rho_{\phi}} =   \frac{\frac{1}{2}\dot{\phi}^2 - V(\phi)}{\frac{1}{2}\dot{\phi}^2 + V(\phi)} =
\begin{cases}
    \geq 1, & V(\phi) \leq 0, \cr
     \leq 1, & V(\phi) \geq 0, \cr
\end{cases}, ~~~
\eqn
provided that $\rho_{\phi} > 0$.

Eqs.(\ref{eq2.6}), (\ref{eq2.7}),  (\ref{eq2.14}) and (\ref{eq2.15}) are the first-order ordinary differential equations for the four canonical variables ($v, b; \phi, p_{\phi}$). Once the initial conditions are specified at a given moment,  they uniquely determine the trajectory of the evolution of the Universe. Such initial conditions are often imposed at the quantum bounce \cite{Ashtekar:2011ni,Li:2021mop}, at which the expansion factor reaches its minimal value and the energy density reaches its maximum.

It should be noted that these four first-order dynamical differential equations are equivalent to the two second-order differential equations given by Eqs.(\ref{eq2.10}) and (\ref{eq2.16}).

In addition, the advantage of imposing the initial conditions at the bounce is that the time derivative of the scalar field at the bounce $\dot\phi_B$ is determined uniquely up to a sign for any given initial scalar field value at the bounce $\phi_B$ via the relation $\rho(t_B) = \rho_c$, where $t_B$ denotes the time of the bounce,  
which yields
\bq
\lb{eq2.19}
\dot\phi_B = \pm\sqrt{2(\rho_c-V(\phi_B))}.
\eq
On the other hand, from Eqs.(\ref{eq2.10}) and (\ref{eq2.16})  
we can see that these equations are scaling-invariant with respect to the expansion factor $a \rightarrow a/L_o$. Therefore, without loss of generality, we can always set the scale factor at the bounce $a_B = 1$, which is equivalent to setting $v_B = 1$. Then, the initial conditions are reduced to the choice of
\bq
\lb{eq2.20}
\left(\phi_B, \text{sgn}\left(\dot\phi_B\right)\right).
\eq
Moreover, using the translation invariance $t \rightarrow t + t_0$, in the rest of this paper, we shall set $t_B = 0$.

%%%%%%%%%%%%%%%%%%%%%%%%%%%%%%%%%%%%%%%%%%%%%%%%%%%%%%%%%%

\subsection{Effective Dynamical Equations in mLQC-I}

In  the framework of mLQC-I, the dynamics can be obtained directly from the effective Hamiltonian \cite{Yang:2009fp,Li:2018fco}
\bq
\lb{Hamiltonian for mLQC-I}
\mathcal{H} = \frac{3v}{8\pi G \lambda^2 } \left\{\sin^2 (\lambda b) -\frac{(\gamma^2+1)\sin^2 (2\lambda b)}{4\gamma^2}\right\}+\mathcal{H_M}.
\eq 
Then, for a scalar field with its Hamiltonian given above, the physical variables $b$ and $v$ satisfy the following Hamiltonian equations
\bqn
\lb{eqA}
\dot v &=& \{v,\mathcal{H}\} \nb\\
&=& \frac{3v\sin{(2\lambda b)}}{2\gamma \lambda}\left\{(\gamma^2+1)\cos{(2\lambda b)}-\gamma^2\right\}, \\
\lb{eqB}
\dot b &=& \{b,\mathcal{H}\}
= \frac{3\sin^2{(\lambda b)}}{2\gamma\lambda^2}\left\{\gamma^2\sin^2{(\lambda b )}-\cos^2{(\lambda b )}\right\} \nb\\
&& ~~~~~~~~~~~~~ - 4\pi G \gamma P_{\phi}, 
\eqn
while the equations for $\phi$ and $p_{\phi}$ take the same forms as those given by Eqs.(\ref{eq2.14}) and 
(\ref{eq2.15}).

Similar to the LQC case, the above Hamiltonian equations can be also cast in the modified Friedman-Raychaudhuri (FR) forms \cite{Li:2021mop}
\begin{widetext}
\bqn
\lb{FRa}
H^2 &=& \frac{8\pi G \rho}{3} \left(1-\frac{\rho}{\rcone} \right) \left(1+\frac{\gamma^2 \rho/\rcone}{\left(\gamma^2+1\right)\left(1+\sqrt{1-\rho/\rcone}\right)^2}\right), \;\;\; (t \ge t_B), \\
\lb{FRb}
\frac{\ddot{a}}{a} &=& -\frac{4 \pi G}{3}\left(\rho+3P\right) + \frac{4\pi G \rho^2}{3 \rcone}\left(\frac{\left(7\gamma^2+8\right)-4\rho/\rcone\left(5\gamma^2+8\right)\sqrt{1-\rho/\rcone}}{\left(\gamma^2+1\right)\left(1+\sqrt{1-\rho/\rcone}\right)^2}\right)  \nb\\
&& +4\pi G P\left(\frac{3\gamma^2+2+2\sqrt{1-\rho/\rcone}}{\left(\gamma^2+1\right)\left(1+\sqrt{1-\rho/\rcone}\right)}\right)\frac{\rho}{\rcone}, \;\;\; (t \ge t_B),
\eqn
\end{widetext}
where 
\bqn
\lb{eq_rho}
\rcone \equiv \frac{\rho_c}{4(1+\gamma^2)}.  
\eqn
From Eqs.(\ref{FRa}) and (\ref{FRb}) it can be shown that the energy conservation law 
\bq
\lb{eq2.27aa}
\dot\rho+3H(\rho+P) = 0,
\eq
holds.
Substituting  Eq.(\ref{eq2.8}) into it, we find that it also yields the same Klein-Gordon equation (\ref{eq2.16}), 
while in terms of $\rho$ and $P$, we find that $\dot{b}$ is also given by Eq.(\ref{eq2.9}).

It should be noted that,   Eqs.(\ref{FRa}) and (\ref{FRb}) hold only after the quantum bounce ($t \ge t_B$), as already indicated in these equations, at which we have $\rho(t_B) = \rcone$ and $H(t_B) = 0$, so the expansion factor reaches its minimal value $a_B \equiv a(t_B)$. When $t \gg t_B$ (or equivalently, $\rho/\rcone \ll 1$), Eqs.(\ref{FRa}) and (\ref{FRb}) reduce to their relativistic limits
\bqn
\lb{FRc1}
H^2 &\simeq& \frac{8\pi G}{3} \rho,\; (t \gg t_B), \\
\lb{FRc2}
\frac{\ddot{a}}{a} &\simeq&  - \frac{4\pi G}{3}\left(\rho + 3P\right), \; (t \gg t_B).
\eqn
In particular, it is interesting to note that $\rho/\rcone \simeq 10^{-12}$ at the onset of inflation \cite{Ashtekar:2011ni,Li:2021mop}.
Therefore, during the inflationary phase, the modified FR equations are well approximated by its classical limits (\ref{FRc1}) and (\ref{FRc2}).

In the pre-bounce phase ($t \le t_B$), the modified FR equations take the form \cite{Li:2021mop}
\begin{widetext}
\bqn
\lb{FRaa}
H^2 &=& \frac{8\pi G_{\alpha} \rho_\Lambda}{3}\left(1-\frac{\rho}{\rcone}\right) \left(1+\frac{\rho\left(1-2\gamma^2+\sqrt{1-\rho/\rcone}\right)}{4\gamma^2\rcone\left(1+\sqrt{1-\rho/\rcone}\right)}\right), \;\;\; (t \le t_B), \\
\lb{FRbb}
\frac{\ddot{a}}{a} &=& -\frac{4\pi G_{\alpha}}{3}\left(\rho+3P-2\rho_\Lambda\right)
+4\pi G_{\alpha} P \left(\frac{2-3\gamma^2+2\sqrt{1-\rho/\rcone}}{\left(1-5\gamma^2\right)\left(1+\sqrt{1-\rho/\rcone}\right)}\right)\frac{\rho}{\rcone} \nb\\
&&
-\frac{4\pi G_{\alpha} \rho^2\left(2\gamma^2 + 5\gamma^2\left(1+ \sqrt{1-\rho/\rcone}\right)  - 4 \left(1+\sqrt{1-\rho/\rcone}\right)^2\right)}{3\rcone\left(1-5\gamma^2\right)\left(1+\sqrt{1-\rho/\rcone}\right)^2}, \;\;\; (t \le t_B),
\eqn
\end{widetext}
where $G_{\alpha} \equiv \alpha G$, and
\bq\lb{eq2.18}
\alpha \equiv \frac{1-5\gamma^2}{\gamma^2+1}, \quad
\rho_\Lambda \equiv \frac{3}{8\pi G_{\alpha}\lambda^2(1+\gamma^2)^2}.
\eq
From Eqs.(\ref{FRaa}) and (\ref{FRbb}) we can see that at the bounce $\rho(t_B) = \rcone$, the universe contracts to its minimal volume $v = a_B^3$ at $t = t_B$.
Afterward, it smoothly passes to the expansion phase, but is now described by Eqs.(\ref{FRa}) and (\ref{FRb}). The smoothness is shown explicitly in \cite{Li:2018opr,Li:2018fco,Li:2019ipm}, and can be also seen from  Eqs.(\ref{eq2.12}) and (\ref{eq2.13}), which hold across the bounce.

When $t \ll t_B$ (or $\rho/\rcone \ll 1$),  Eqs.(\ref{FRaa}) and (\ref{FRbb}) reduce to
\bqn
\lb{FRccA}
H^2 &\simeq& \frac{8\pi G_{\alpha}}{3} \rho_{\Lambda} \left(1 - \frac{\rho}{\rho_{\Lambda}}\right),
\; (t \ll t_B),\\
\lb{FRccB}
\frac{\ddot{a}}{a} &\simeq&   \frac{8\pi G_{\alpha}}{3}\rho_{\Lambda}\left( 1 - \frac{\rho + 3P}{2\rho_{\Lambda}}\right), \; (t \ll t_B), ~~~~
\eqn which are quite different from Eqs.(\ref{FRc1}) and (\ref{FRc2}). In particular, the effective Planck-scale cosmological constant $\rho_{\Lambda}$ soon dominates the evolution of the pre-bounce phase, whereby a de Sitter spacetime is obtained in the pre-bounce phase but with a Planck-scale cosmological constant $\rho_{\Lambda} \simeq {\cal{O}}(\rho_{\text{pl}})$. In addition, the Newtonian constant $G$ is replaced by $G_{\alpha} (= \alpha G)$, where $\alpha$
is defined by Eq.(\ref{eq2.18}). More remarkably, this Planck-scale cosmological constant is filtered out by the quantum bounce and disappears miraculously after the bounce, whereby the classical FR equations are obtained, as shown explicitly by Eqs.(\ref{FRc1}) and (\ref{FRc2}). This is significantly different from LQC \cite{Ashtekar:2011ni}, in which the evolution of the Universe is symmetric with respect to the bounce \footnote{More precisely, it is symmetric for kinetic energy-dominated initial conditions $\dot{\phi}_B^2 \gg 2V(\phi_B)$ \cite{Ashtekar:2011ni,Li:2021mop}.}.

With similar arguments as those given in LQC, the initial conditions of the dynamical system of  Eqs.(\ref{eqA}), (\ref{eqB}), (\ref{eq2.14}) and 
(\ref{eq2.15}) also reduce to Eq.(\ref{eq2.20}) but now with 
\bq
\lb{eq2.36}
\dot\phi_B = \pm\sqrt{2(\rho_c^{\text{I}}-V(\phi_B))}.
\eq

\section{Effects of Ekpyrotic Mechanism on Inflation in LQC}
\renewcommand{\theequation}{3.\arabic{equation}}
\setcounter{equation}{0}
\lb{SecIII}

It is well-known that shear behave like a stiff fluid  \cite{Ryan:1975jw}
\bq
\lb{eq3.1}
\sigma^2 \equiv \sigma_{\mu\nu}\sigma^{\mu\nu} = \frac{\Sigma^2}{a^6},
\eq
where $a(t)$ is the average expansion factor of the 3-volume of a universe, $\Sigma^2$ is a constant, and $\sigma_{\mu\nu}$ denotes the anisotropic shear tensor, defined via the relation
\bqn
\lb{eq3.1aa}
\nabla_{\nu}v_{\mu} = \frac{1}{3}\left(g_{\mu\nu} + v_{\mu}v_{\nu}\right)\theta + \omega_{\mu\nu} + \sigma_{\mu\nu}.
\eqn
Here $v^{\mu}$ denotes the unit tangential vector of the time-like geodesics, $\theta$ and $\omega_{\mu\nu}$  denote respectively the expansion scalar and vorticity  tensor of the time-like geodesics. In the homogeneous universe, we have  $\omega_{\mu\nu} = 0$ and $\theta = 3H$.   

 For the kinetic energy dominated initial conditions, the scalar field also behaves like a stiff fluid, so we have 
\bq
\lb{eq3.2}
\rho_{\phi} \simeq P_{\phi} = \frac{\rho^{(0)}_{\phi}}{a^6},\;\; (t \simeq t_B),
\eq
where   $\rho^{(0)}_{\phi}$ is a constant. 
Therefore, it is not always clear which one shall dominate the evolution of the universe near the bounce. If the shear dominates, the universe will become highly anisotropic after the bounce, whereby a homogeneous and isotropic universe cannot be developed. Therefore, {\em it is crucial for any bounce model, including LQC and mLQC-I, to be considered as viable, one has to to make sure that the shear does not dominate in the contracting phase, especially near the bounce} \cite{Lehners:2008vx,Battefeld:2014uga,Brandenberger:2016vhg}. 
One way   is to introduce the ekpyrotic potential \cite{Khoury:2001wf,Cai:2012va,Motaharfar:2023hil}
 \begin{equation}
 \lb{eq3.3}
    V_{\text{ekp}}(\phi)=-\frac{2 U_0}{e^{-\sqrt{\frac{16 \pi}{p}}\phi}+e^{\beta\sqrt{\frac{16 \pi}{p}}\phi}},
\end{equation}
so that near the bounce we have $V(\phi_B) < 0$, where $U_0, \; p$ and $\beta$ are all positive and otherwise free parameters. 
Then, we have
$w_{\phi} > 1$, and
\begin{equation}
\lb{eq3.4}
\rho_{\phi}^{\text{ekp}}   \propto \frac{1}{a^{3(1+w_{\phi})}},\;\; (t \simeq t_B),
\end{equation}
so the scalar field will dominate the evolution of the universe and the effects of the shear will be suppressed. As a result, the contracting universe can smoothly evolve into an expanding  homogeneous and isotropic one. 

When far away from the bounce, we would expect to obtain an inflationary phase in the post-bounce region, $t \gg t_B$ \footnote{It should be noted that in most of the bouncing models,  the inflationary phase is not required, see, for example,   \cite{Lehners:2008vx,Battefeld:2014uga,Brandenberger:2016vhg}. This is fundamentally different from quantum bouncing models of LQG, in which it has been shown that inflation after the bounce is generic in LQC \cite{Ashtekar:2011rm} amd mLQCs \cite{Li:2019ipm}.}. This is possible if the total potential $V(\phi)$ consists of two parts
\begin{equation}
 \label{eq3.5}
    V(\phi)=V_\text{ekp}(\phi)+V_{\text{inf}}(\phi),
\end{equation}
where   $V_{\text{inf}}(\phi)$ denotes an inflationary potential and will dominate the evolution of the universe when $t \gg t_B$, while for $t \simeq t_B$ the ekpyrotic potential $V_{\text{ekp}}(\phi)$  dominates. 

Following Planck 2018 data \cite{BICEP:2021xfz}, inflation with various known potentials have been ruled out, including   potentials with the form $V(\phi) \propto \phi^n$.  However, models with polynomial chaotic potentials can fit the observations well \cite{Destri:2007pv,Nakayama:2013jka,Kallosh:2014xwa}. A typical example is \cite{Kallosh:2025ijd}
\begin{equation}
\label{eq3.11}
V_{\text{inf}}(\phi) = \frac{1}{2}m^2\phi^2\left(1 - \alpha_1\phi + \alpha_2\phi^2\right)^2,  
\end{equation}
where $\alpha_{1, 2}$ are two coupling constants. By properly choosing these constants, it can be shown that the models
fit the observational data very well. In particular, choosing $\alpha_1 = 0.14$ and $\alpha_2 = 6.644\times 10^{-3}$
allows the model to fit very well to the current Atacama Cosmology Telescope (ACT) observations \cite{ACT:2025fju}. 
In this paper, we shall consider the polynomial chaotic potentials given above as a representative case, and the generalization of our analysis to other viable potentials are straightforwards.

Then, a natural question is whether or not a mechanism mentioned above exists. Our following  analysis shows that this can indeed be the case by properly choosing the parameters involved in the models, despite the fact that the effects of the ekpyrotic-like potential are dramatic. 

\begin{figure}[h!]
\includegraphics[width=0.85\linewidth]{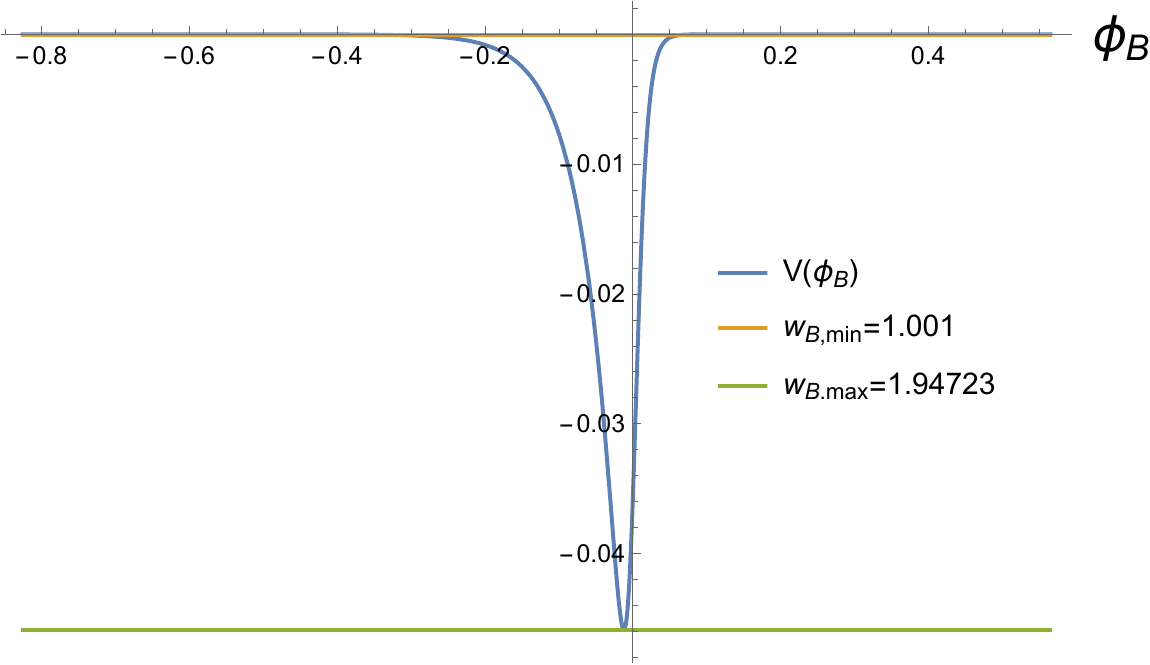}
\caption{The plot of the total potential $V(\phi)$ defined by Eq.(\ref{eq3.5}) with $U_0=0.0366$, $p=0.1$, $\beta=5$ and the chaotic inflationary potential given by Eq.(\ref{eq3.11}) with $\alpha_1 = \alpha_2 = 0,\; m = 1.26 \times 10^{-6}\; m_{P}$. The corresponding  minimal and maximal values of $w_B$ are also given. Here $m_p \equiv 1/(8\pi G)^{1/2}$.}
\label{V_vs_phi_b-0.0366-0.1-5}
\end{figure}

For our above claim, let us first show how to choose the initial conditions at the bounce $t = t_B$ with a total potential given by Eq.(\ref{eq3.5}). First, from Eqs. (\ref{eq2.18}), (\ref{eq2.19}) and (\ref{eq2.36}) we find  
\begin{equation}
    \label{eq3.6}
    V(\phi_B)= - \frac{w_B - 1}{2}\rho_B,
\end{equation}
where $\rho_B = (\rho_c, \rho_c^{\text{I}})$, depending on whether we are working in the framework of LQC or mLQC-I. For any given potential $V(\phi)$ and a fixed equation of state $w_B > 1$, we can solve the above equation for $\phi_B$.  In particular, starting with a minimal value of $w_B$, say, $w_{\text{Bmin}}$,   we can solve Eq.(\ref{eq3.6}) numerically to obtain  the corresponding values of $\phi_B$. As shown in Fig.\ \ref{V_vs_phi_b-0.0366-0.1-5}, the maximal value of 
 $w_{\text{Bmax}}$ is obtained when the potential is at its minimum  $V_{\text{min}}(\phi_B)$ with
\begin{equation}
    \label{eq3.6a}
    V_{\text{min}}(\phi_{B}) = - \frac{w_{\text{Bmax}} - 1}{2}\rho_B,
\end{equation} 
denoted by the crossing point of the horizontal straight line $- (w_{\text{Bmax}} - 1)\rho_B/2$ and $V(\phi_B)$.

% Figure 1 %%%%%%%%%%%%%%%%%%%%%%%%%%%%%%%%%%%%%%%%%%%%%%%%%%%%%%%%%%%%%%%%%%%%%%%%%%%%%%%%%%%%%%%%%%%%%%%%%%%%%
%%%%%%%%%%%%%%%%%%%%%%%%%%%%%%%%%%%%%%%%%%%%%%%%%%%%%%%%%%%%%%%%%%%%%%%%%%%%%%%%%%%%%%%%%%%%%%%%%%%%%%%%%%%%%%%%
%Figure1%%%%%%%%%%%%%%%%%%%%%%%%%%%%%%%%%%%%%%%%%%%%%%%%%%%%%%%%%%%%%%%%%%%%%%%%%%% 

With the above chosen initial conditions for $\phi_B$, we can study the evolution of the universe for any given   inflationary potential. Before doing so, let us first introduce some relevant quantities. 

\begin{itemize}

\item The first-order Hubble rate and potential slow-roll parameters \cite{Baumann:2009ds}
\begin{align}
\lb{eq3.11a}
\epsilon_H &= -\frac{\dot H}{H^2}, &\eta_H =  \frac{\ddot H}{2H \dot H}, \\
\lb{eq3.11b}
\epsilon_V &= \frac{1}{16\pi G}\left( \frac{V_{,\phi}}{V}\right)^2,  &\eta_V = \frac{V_{,\phi\phi}}{8\pi G V}.
\end{align}
These sets of slow-roll parameters are typically used for different purposes. In particular, the slow-roll parameters with the subscript ``$V$" can be used to determine which part of the potential can successfully drive inflation. On the other hand, slow-roll parameters with subscript ``$H$" are used for numerical simulations to define when slow-roll inflation begins and ends. Kinematically, we have
\begin{equation}\label{a and H relation}
\frac{\ddot a}{a} = H^2(1-\epsilon_H).
\end{equation}
The Universe experiences an accelerated expansion whenever $\epsilon_H < 1$, whereas slow-roll inflation occurs only when \cite{Baumann:2009ds}
\bqn
\lb{eq3.14a}
\epsilon_H(t), \;\; \left|\eta_H(t)\right| \ll 1. 
\eqn
For the sake of concreteness, we define the onset of inflation as the time $t_i$ when $\epsilon_H(t_i) = 1$ for the first time in the transition phase, where $\epsilon_H < 1$ for $t > t_i$. The end of the inflationary phase is defined at the time $t_{\text{end}}$ when $\epsilon_H(t_{\text{end}}) = 1$ again for the first time after $t_i$. Therefore, for $t \in (t_i, t_{\text{end}})$ we have $\epsilon_H < 1$ and $\ddot{a} > 0$, that is, the universe is in its inflationary phase.  

Certainly, for the inflationary phase to be slowly rolling, 
the conditions (\ref{eq3.14a}) need to be satisfied during the inflation. Once these conditions are satisfied, we have \cite{Yogesh:2024iip}
\bqn
\lb{eq3.14b}
\epsilon_H \simeq \epsilon_V, \quad \eta_H \simeq \eta_V - \epsilon_V, \;\; (\epsilon_H, |\eta_H| \ll 1). 
\eqn

\item The e-fold $N_{\text{inf}}$ during the inflationary phase is  defined as
\begin{equation}\label{efolds definition}
N_{\text{inf}}  = \ln\left({\frac{a(t_{\text{end}})}{a(t_i)}}\right).
\end{equation}
To have a successful inflation, the inflation potential has to be very flat, so that the Universe can expand large enough \cite{Baumann:2009ds}. All the cosmological problems can be resolved if the Universe expands about 60 e-folds during the inflationary phase, although its exact value depends on the inflationary models \cite{Planck:2018jri}. Therefore, 
in the following we shall require   $N_{\text{inf}} \gtrsim 60$, although our main conclusions do not depend on its precise value.  From the above definition it is clear that in general $N_{\text{inf}}$ depends on the specific value of $\phi_B$.

\end{itemize}

With the above in mind, we are now ready to solve the dynamical equations respectively in LQC and mLQC-I given in the last section for a given inflationary potential  $V_{\text{inf}}(\phi)$.   In the rest of this section, we shall focus on LQC, while leaving the studies of mLQC-I to the next section.

\subsection{Inflation Without the Ekpyrotic Mechanism}

 To see the effects of the Ekpyrotic mechanism on the inflation phase, let us first briefly review the probability  of inflation in LQC  without Ekpyrotic Mechanism \cite{Ashtekar:2011rm}.
 In doing so, we find that it is simple and instructive to start with the chaotic potential  $\alpha_1 = \alpha_2 = 0$, despite the fact that this potential has been already ruled out by observations \cite{Planck:2018jri}. This is due to the fact that our main conclusions do not depend on the specific forms of the potentials. Then, we shall turn to the potentials with $\alpha_1 \alpha_2 \not= 0$, which are favorable to observations, and find that indeed similar effects occur.

Without the Ekpyrotic mechanism, since $\rho \leq \rho_c$, we find that $\phi$ must satisfy the condition $\dot\phi^2/2 + V(\phi) \leq \rho_c$. In particular, the equation 
\bqn
\lb{phiMN}
V_{\text{inf}}(\phi) = \rho_c,
\eqn
determines the range of the scalar field $\phi$. For example, for the chaotic potential we find that 
\bqn
\lb{eq3.16bb}
\phi \in \left(\phi_{\text{min}}, \phi_{\text{max}}\right), 
\eqn
where $\phi_{\text{min}} = - \phi_{\text{max}} = - \sqrt{2\rho_c/m^2}$.

%%%%%%%%%%%%%%%%
%% Table I
%%%%%%%%%%%%%%%%
\begin{table*}[t!]
\centering
\resizebox{\textwidth}{!}{%
\begin{tabular}{|c|c|c|r|c|c|r|r|c|}
\hline
                          & $m$                  & $H_*$                & \multicolumn{1}{c|}{$\phi_*$} & $\phi_{\text{min}}$ & $\phi_{\text{max}}$ & \multicolumn{1}{c|}{$\phi_1$} & \multicolumn{1}{c|}{$\phi_2$} & $P_{\text{LQC}}$     \\ \hline
\multirow{2}{*}{Planck}   & $1.56\times 10^{-6}$ & $5.54\times 10^{-6}$ & $2.52$                        & $-436.64$           & $450.69$            & $-5.57$                       & $0.33$                        & $7.30\times 10^{-3}$ \\ \cline{2-9} 
                          & $6.93\times 10^{-7}$ & $8.73\times 10^{-6}$ & $-3.79$                       & $-574.36$           & $588.41$            & $-5.70$                       & $0.20$                        & $5.57\times 10^{-3}$ \\ \hline
\multirow{2}{*}{ACT}      & $1.49\times 10^{-6}$ & $5.37\times 10^{-6}$ & $2.58$                        & $-443.25$           & $457.29$            & $-5.57$                       & $0.33$                        & $7.20\times 10^{-3}$ \\ \cline{2-9} 
                          & $6.51\times 10^{-7}$ & $8.57\times 10^{-6}$ & $-3.90$                       & $-586.66$           & $600.70$            & $-5.70$                       & $0.18$                        & $5.43\times 10^{-3}$ \\ \hline
\multirow{2}{*}{Combined} & $1.11\times 10^{-6}$ & $4.30\times 10^{-6}$ & $2.92$                        & $-489.64$           & $503.68$            & $-5.63$                       & $0.28$                        & $6.52\times 10^{-3}$ \\ \cline{2-9} 
                          & $4.31\times 10^{-7}$ & $7.51\times 10^{-6}$ & $-4.70$                       & $-673.77$           & $687.82$            & $-5.78$                       & $0.13$                        & $4.76\times 10^{-3}$ \\ \hline
\end{tabular}%
}
\caption{The physical quantities ($H_*, \phi_*, m$) for a given set of ($A_s, n_s$) from the observations, Planck, ACT or Planck + ACT + LB2 (Combined),  and the corresponding values of $\phi_{\text{min}}$, $\phi_{\text{max}}$, $\phi_1$, $\phi_2$, and  $P_{\text{LQC}}$ for   $\dot{\phi}_B > 0$, and $\alpha_1 = 0.14$ and $\alpha_2 = 6.644\times 10^{-3}$  
\cite{Kallosh:2025ijd} in LQC.}
\lb{TableI}
\label{LQCpolyTab}
\end{table*}
 
%%%%%%%%%%%%%%%%
%% Fig.\ 2
%%%%%%%%%%%%%%%%
\begin{figure*}[t!] 
    \centering
    \includegraphics[width=7cm]{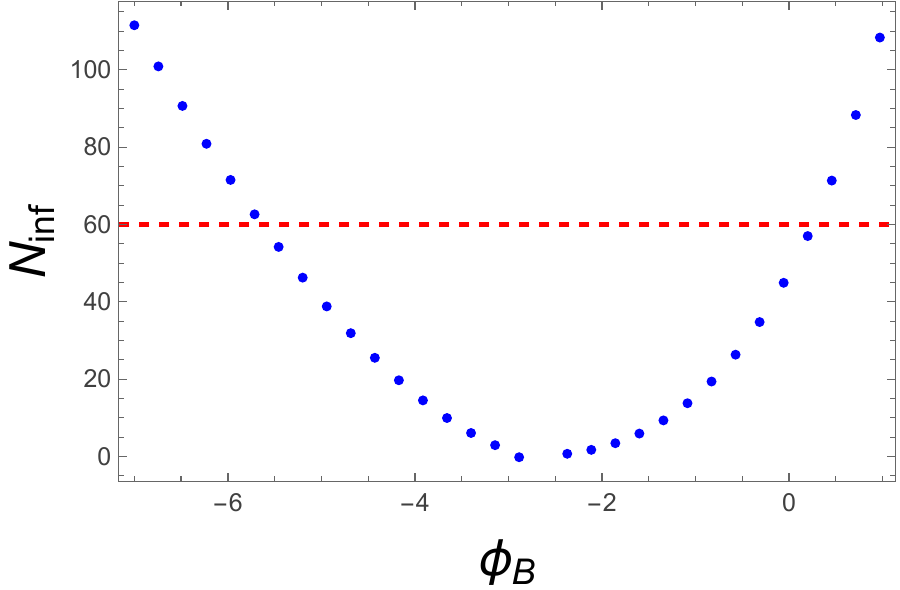} 
    \includegraphics[width=7cm]{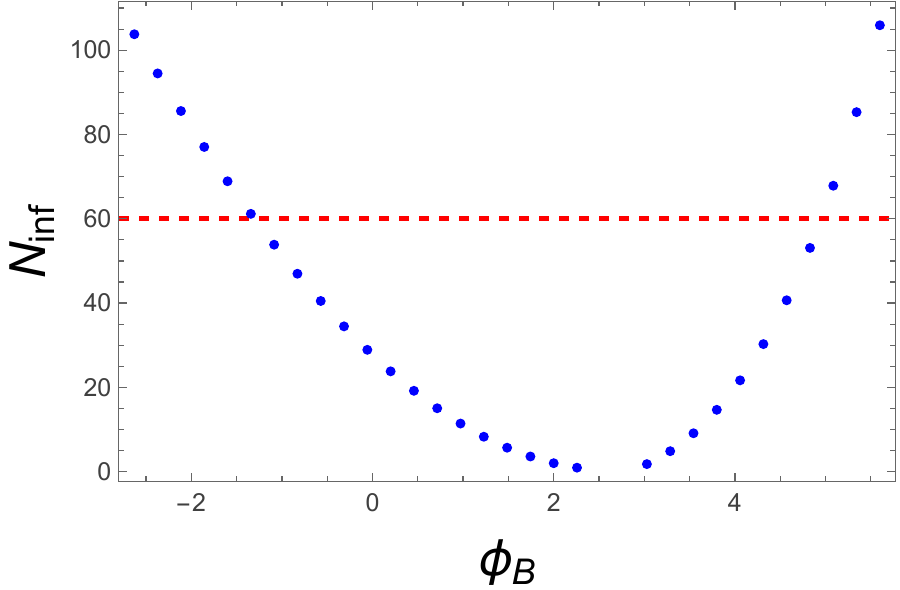} \\  
    \vspace{.2cm}
   \mbox{\hspace{1.cm}}  (a) \mbox{\hspace{6.8cm}} (b)\\
    \caption{The plots of the total number of e-folds  during the inflationary phase in LQC, when the  polynomial chaotic potential given by Eq.(\ref{eq3.11}) is  present with the choice $\alpha_1 = 0.14, \; \alpha_2 = 6.644 \times 10^{-3}$ and $m = 1.11 \times 10^{-6}\; m_\text{pl}$. Plots (a) and (b) are respectively for $\dot\phi_B >0$ and  $\dot\phi_B < 0$.}
    \label{polyefolds}  
\end{figure*}

The probability $P(E)$ that an event $E$ will occur is given by \cite{Ashtekar:2011rm}
\bq
\lb{4.9}
P(E) = \frac{1}{{\cal{D}}} \int_{\mathcal{I}(E)}{d\omega},
\eq
where 
\bqn
\lb{4.9aa}
d\omega &\equiv&  \left\{-2\left[\hat{\cal{H}}^{\text{grav}}(b_B)   +  V(\phi)\right]\right\}^{1/2} d\phi,\nb\\
\hat{\cal{H}}^{\text{grav}} &\equiv& \frac{1}{v} {\cal{H}}^{\text{grav}}
= \frac{1}{v} \left({\cal{H}} -  {\cal{H}}_{\cal{M}}\right), 
\eqn
$\mathcal{I}(E)$ is the interval on the $\phi_B$-axis, which corresponds to the physically distinct initial conditions in which the event $E$ happens, and ${\cal{D}}$ is the total measure 
\bq
\lb{4.10}
 {\cal{D}} \equiv  \int_{\phi_{\mathrm{min}}}^{\phi_{\mathrm{max}}} {d\omega}.
\eq 
To have $N_{\text{inf}} \gtrsim 60$, it was found that $\phi_B$ must fall into the windows \cite{Ashtekar:2011rm}
\bq
\lb{4.10bb}
\phi_B \in \left(-\phi_{\text{max}}, -5.5 m_P\right) \cup \left(0.94 m_P, \phi_{\text{max}}\right),
\eq
where from Eq.(\ref{phiMN}) we find $\phi_{\text{max}} \equiv \sqrt{2\rho_c/m^2} \simeq 7.478 \times 10^5\; m_P$
for the chaotic potential of Eq.(\ref{eq3.11}) with $\alpha_1 = \alpha_2 = 0, \; m = 1.21\times 10^{-6} m_P$ \footnote{In \cite{Ashtekar:2011rm} the mass parameter $m$ was calculated based on the WMAP observation \cite{2011ApJS..192...18K}, $A_s = 2.43\times 10^{-9}, \; n_s = 0.968$ 
where $A_s$ and $n_s$ represent the amplitude and spectral index of the pivot mode $k_*/a_0 = 0.002/\text{Mpc}$ of the scalar perturbations, respectively, and $a_0$ is the current value of the expansion factor of the Universe.}. Then, the probability that the desired slow-roll inflation cannot be realized is given by
\bqn
\lb{4.10cc}
P_{\text{LQC}} \left(\text{not realized}\right) \lesssim \frac{\int_{-5.5 m_P}^{0.94 m_P}{d\omega}}
{\int_{-\phi_{\text{max}}}^{\phi_{\mathrm{max}}} {d\omega}}
\lesssim
5.48\times 10^{-6}. \nb\\
\eqn

When $\alpha_1\alpha_2 \not= 0$, following \cite{Ashtekar:2011rm,Yogesh:2024iip}, we need first to calculate the mass parameter $m$  for a given set of $(A_s(k_*), n_s(k_*))$ from observations, where quantities with the subscript ``*" represent the ones calculated at the horizon-crossing $k = aH$ for the pivot mode $k_*/a_0$, where $a_0$ denotes the current value of the expansion factor of the Universe, and
$A_s$ and $n_s$ are respectively  the amplitude and spectral index of the scalar perturbations. 
With the slow-roll conditions (\ref{eq3.14a}), the following equations hold at  the horizon-crossing \cite{Yogesh:2024iip}
\bqn
\lb{eqA.38}
 && H_*^2 = \pi m_{\text{P}}^2 \frac{\epsilon_V^* A_s(k_*)}{{\cal{F}}(k_*)},\\
 \lb{eqA.39}
 &&  \xi_{\text{nBD}}(k_*)  - 2\left(3\epsilon_V^* - \eta_V^*\right) = n_s(k_*) -1, \\
 \lb{eqA.40}
&& H_*^2 \simeq   \frac{8\pi }{3 m_{\text{P}}^2 }V(V_0,\phi_*),
\eqn
where \cite{Zhu:2017jew}
\begin{widetext}
\bqn
\lb{eqA.43}
 {\cal{F}}(k) &=& 1 + \delta_{\text{PL}}(k) =  1+ \left[1 + \cos\left(\frac{\pi}{\sqrt{3}}\right)\right]{\text{csch}}^2 \left(\frac{\pi k}{\sqrt{6} k_B}\right),\quad k_B \equiv \sqrt{\frac{8\pi \rho_c}{m_{P}^2}},
 \eqn
 \end{widetext}
 and
 \bqn
 \lb{eqA.27}
 \xi_{\text{nBD}}(k_*)  &\equiv&  \left. \frac{d\ln{\cal{F}}(k)}{d\ln k}\right|_{k = k_*}.
\eqn
In addition to the above three equations, there are three more independent ones  given by \cite{Yogesh:2024iip}
\bqn
  \lb{eqA.41}
 && k_* = a_* H_*,\\
 \lb{eqA.42}
&& \dot\phi_* \simeq - \frac{V_{,\phi}(V_0,\phi_*)}{3H_*},\\
 \lb{eqA.43}
 && a_B(a_*, \phi_*, \dot\phi_*) =  1. 
\eqn
Therefore, at the horizon-crossing there are six independent algebraic equations, which uniquely determine the six unknown parameters, $\left(a_*, H_*, \phi_*, \dot\phi_*, k_*, V_0\right)$ for a given set of $\left(A_s, n_s\right)$ from observations. It is remarkable to note that once $\left(a_*, \phi_*, \dot\phi_*\right)$ are known, the modified Friedmann and Klein-Gordon equations will uniquely determine the entire evolution of the Universe, from the pre-bounce phase $t < t_B$ to the post-bounce phase $t > t_B$ \cite{Yogesh:2024iip}.

Considering the fact that $k_* = a_0 e^{N_T} \simeq {\cal{O}}(m_P)$, where $N_T (\equiv \ln(a_0/a_B) \gtrsim 141)$ \cite{Zhu:2017jew} denotes the total e-fold from the quantum bounce to the current time, it was found that $\left| {\cal{F}}(k_*) - 1\right|, \; \left|\xi_{\text{nBD}}(k_*)\right| \lesssim 10^{-5}$ \cite{Yogesh:2024iip}. Therefore, as the first-order approximation, we  can safely set
${\cal{F}}(k) = 1,\; \xi_{\text{nBD}}(k_*) = 0$. Then, Eqs.(\ref{eqA.38})-(\ref{eqA.40}) reduce to \cite{Ashtekar:2011rm}
\bqn
\lb{eqA.38aa}
 && H_*^2 = \pi m_{\text{P}}^2 \epsilon_V^* A_s(k_*),\\
 \lb{eqA.39aa}
 &&   2\left(3\epsilon_V^* - \eta_V^*\right) = 1 - n_s(k_*), \\
 \lb{eqA.40aa}
&& H_*^2 \simeq   \frac{8\pi }{3 m_{\text{P}}^2 }V(V_0,\phi_*).
\eqn

%%%%%%%%%%%%%%%%
%% Table II
%%%%%%%%%%%%%%%%
\begin{table*}[t!]
\centering
\resizebox{\textwidth}{!}{%
\begin{tabular}{|c|c|c|r|c|c|r|r|c|}
\hline
                          & $m$                  & $H_*$                & \multicolumn{1}{c|}{$\phi_*$} &  $\phi_{\text{min}}$ & $\phi_{\text{max}}$ & \multicolumn{1}{c|}{$\phi_1$} & \multicolumn{1}{c|}{$\phi_2$} & $P_{\text{LQC}}$     \\ \hline
\multirow{2}{*}{Planck}   & $1.55\times 10^{-6}$ & $5.72\times 10^{-6}$ & $2.55$                        & $-458.63$                  & $473.72$            & $-5.55$                       & $0.38$                        & $6.98\times 10^{-3}$ \\ \cline{2-9} 
                          & $7.28\times 10^{-7}$ & $8.68\times 10^{-6}$ & $-3.72$                       & $-592.73$                  & $607.81$            & $-5.68$                       & $0.25$                        & $5.42\times 10^{-3}$ \\ \hline
\multirow{2}{*}{ACT}      & $1.49\times 10^{-6}$ & $5.55\times 10^{-6}$ & $2.60$                        & $-465.47$                  & $480.55$            & $-5.55$                       & $0.35$                        & $6.85\times 10^{-3}$ \\ \cline{2-9} 
                          & $6.85\times 10^{-7}$ & $8.52\times 10^{-6}$ & $-3.83$                       & $-605.11$                  & $620.19$            & $-5.68$                       & $0.23$                        & $5.29\times 10^{-3}$ \\ \hline
\multirow{2}{*}{Combined} & $1.11\times 10^{-6}$ & $4.49\times 10^{-6}$ & $2.95$                        & $-513.27$                  & $528.35$            & $-5.60$                       & $0.30$                        & $6.22\times 10^{-3}$ \\ \cline{2-9} 
                          & $4.59\times 10^{-7}$ & $7.46\times 10^{-6}$ & $-4.62$                       & $-692.69$                  & $707.77$            & $-5.75$                       & $0.18$                        & $4.65\times 10^{-3}$ \\ \hline
\end{tabular}%
}
\caption{The physical quantities ($H_*, \phi_*, m$) for a given set of ($A_s, n_s$) from the observations, Planck, ACT or Planck + ACT + LB2 (Combined),  and the corresponding values of $\phi_{\text{min}}$, $\phi_{\text{max}}$, $\phi_1$, $\phi_2$, and  $P_{\text{LQC}}$ for $\dot{\phi}_B > 0$, and $\alpha_1 = 0.13$ and $\alpha_2 = 5.746\times 10^{-3}$ \cite{Kallosh:2025ijd} in LQC.}
\label{LQCpolyTab2}
\end{table*}

%%%%%%%%%%%%%%%%
%% Table III
%%%%%%%%%%%%%%%%
 \begin{table*}[t!]
\begin{tabular}{|c|c|c|r|r|r|}
\hline
         & $m$                  & $H_*$                & \multicolumn{1}{c|}{$\phi_*$}  & \multicolumn{1}{c|}{$\phi_{\text{Bmin}}$} & \multicolumn{1}{c|}{$\phi_{\text{Bmax}}$} \\ \hline
Planck   & $1.24\times 10^{-6}$ & $7.61\times 10^{-6}$ & $-3.01$                             & $-0.26$                       & $0.05$                        \\ \hline
ACT      & $1.18\times 10^{-6}$ & $7.45\times 10^{-6}$ & $-3.08$                                & $-0.26$                       & $0.05$                        \\ \hline
Combined & $8.80\times 10^{-7}$ & $6.45\times 10^{-6}$ & $-3.58$                                & $-0.26$                       & $0.05$                        \\ \hline
\end{tabular}
\caption{The physical quantities ($H_*, \phi_*, m$) for a given set of ($A_s, n_s$) from the observations, Planck, ACT or Planck + ACT + LB2 (Combined),  and the corresponding values of $\phi_{\text{Bmin}}$, $\phi_{\text{Bmax}}$,   for  $\alpha_1 =\alpha_2 = 0, U_0=0.0366, p=0.1, \beta=5$ and $\dot{\phi}_B > 0$ in LQC.}
\label{LQCchaotTabSingh}
\end{table*}

%%%%%%%%%%%%%%%%
%% Fig.\ 3
%%%%%%%%%%%%%%%%
%%%%%%%%%%%%%%%%%%
  \begin{figure*}[h] 
    \centering    
    \includegraphics[width=7cm]{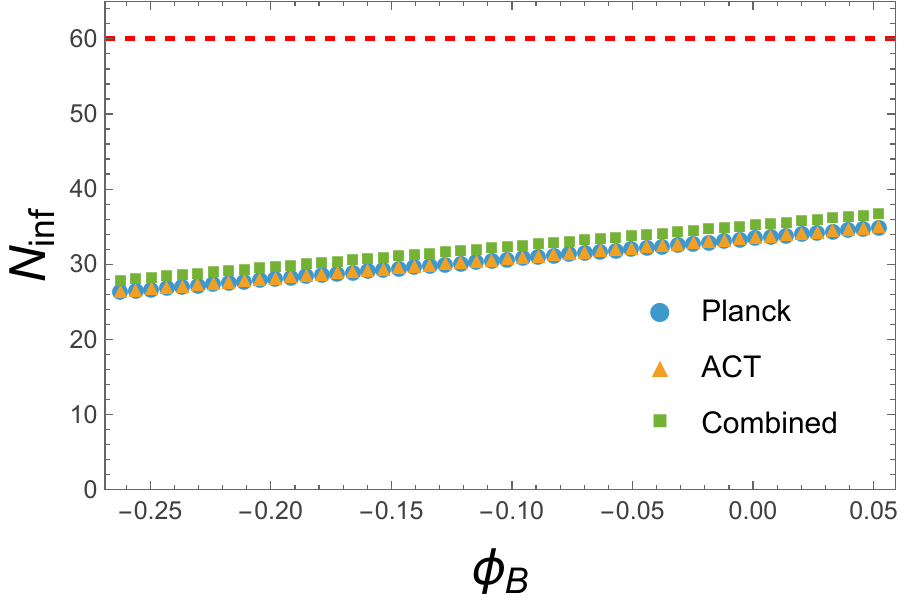} 
     \includegraphics[width=7cm]{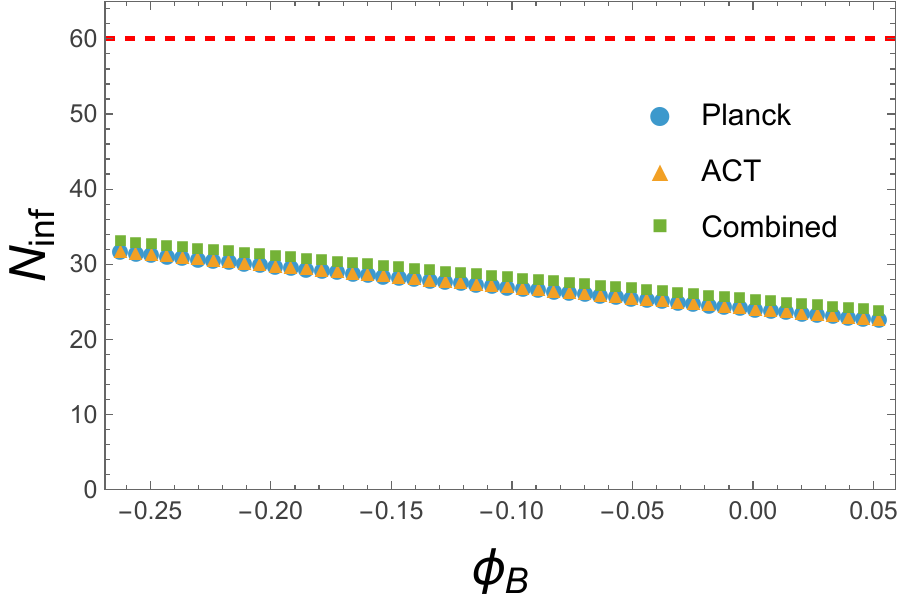}\\  
    %\vspace{.2cm}
   \mbox{\hspace{1.cm}}  (a) \mbox{\hspace{6.8cm}} (b)\\
    \vspace{.5cm}
    \caption{The plots of the total number of e-folds with a chaotic potential, $\alpha_1=\alpha_2=0$, given by Eq.(\ref{eq3.11}) and an ekpyrotic potential given 
    by Eqs.(\ref{eq3.3}) in the framework of LQC, where the parameters are chosen as  $U_0=0.0366$, $p=0.1$, $\beta=5$.  The plots (a) and (b) are  for $\dot\phi_B>0$ and  $\dot\phi_B< 0$, respectively. Three cases are plotted on each graph, corresponding to the three different  masses given in Table \ref{LQCchaotTabSingh}. Note that in the figure the lines for Planck and ACT are practically  indistinguishable.}
    \label{LQCchaotSingh}  
\end{figure*}

%%%%%%%%%%%%%%%%
%% Table IV
%%%%%%%%%%%%%%%%
\begin{table*}[]
\begin{tabular}{|c|c|c|r|r|r|r|r|}
\hline
         & $m$                  & $w_{Bmin}$ & \multicolumn{1}{c|}{$w_{Bmax}$} & \multicolumn{1}{c|}{$\phi_{\text{Bmin}}$} & \multicolumn{1}{c|}{$\phi_{\text{Bmax}}$} & \multicolumn{1}{l|}{$\phi_{60}$} & \multicolumn{1}{l|}{$P_{\text{LQC}}$} \\ \hline
Planck   & $1.23\times 10^{-6}$ & $1.001$    & $1.007$                         & $-0.031$                                  & $0.321$                                   & $0.105$                          & $0.388$                        \\ \hline
ACT      & $1.18\times 10^{-6}$ & $1.001$    & $1.007$                         & $-0.031$                                  & $0.321$                                   & $0.170$                          & $0.572$                        \\ \hline
Combined & $8.80\times 10^{-7}$ & $1.001$    & $1.007$                         & $-0.031$                                  & $0.321$                                   & N/A                              &     $1.0$                        \\ \hline
WMAP     & $1.21\times 10^{-6}$ & $1.001$    & $1.007$                         & $-0.031$                                  & $0.321$                                   & $0.163$                          & $0.551$                        \\ \hline
\end{tabular}
\caption{The physical quantities $m$, $w_{\text{Bmin}}$, $w_{\text{Bmax}}$,  
and $P_{\text{LQC}}(\text{not realized})$  for a given set of ($A_s, n_s$) from the observations, Planck, ACT, Planck + ACT + LB2 (Combined) or WMAP \cite{2011ApJS..192...18K},  and the corresponding values of $\phi_{\text{Bmin}}$ and $\phi_{\text{Bmax}}$   for  $\alpha_1 =\alpha_2 = 0$, $U_0=10^{-3}$, $p=10^{-2}$, $\beta=0.1$ and $\dot{\phi}_B > 0$ in LQC, where $\phi_{60}$ corresponds to the case with $N_{\text{inf}} = 60$.}
\label{LQCchaotTab60}
\end{table*}

%%%%%%%%%%%%%%%%
%% Fig.\ 4
%%%%%%%%%%%%%%%%
%%%%%%%%%%%%%%%%%%
  \begin{figure*}[t] 
    %\centering   
    \includegraphics[width=7cm]{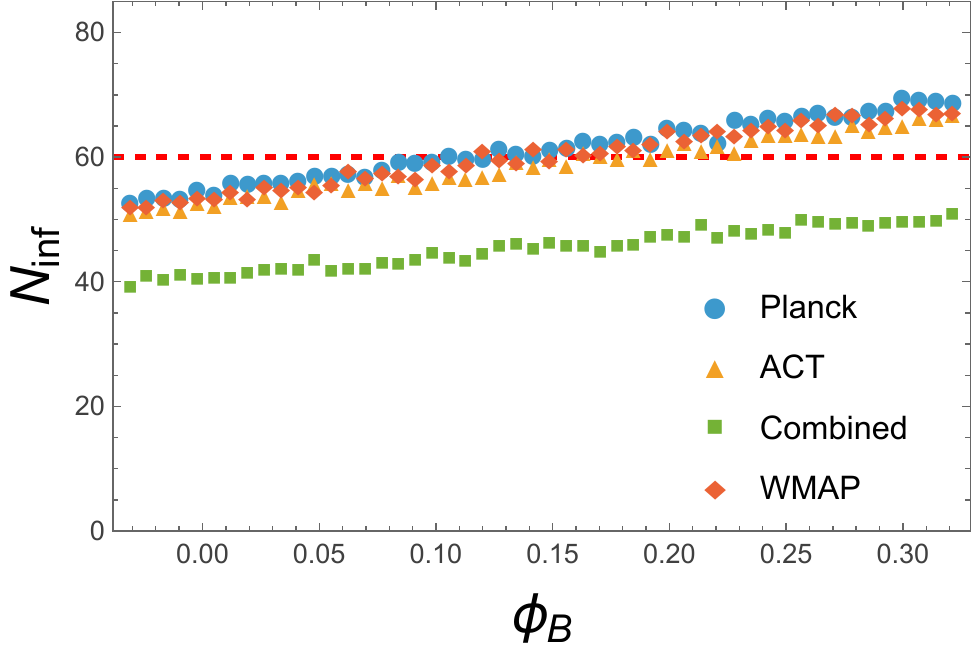} 
    \vspace{.5cm}
    \caption{The plots of the total number of e-folds with a chaotic potential, $\alpha_1=\alpha_2=0$, given by Eq.(\ref{eq3.11}) and an ekpyrotic potential given 
    by Eqs.(\ref{eq3.3}) in the framework of LQC for $\dot{\phi}_B >0$, where the parameters are chosen as  $U_0=10^{-3}$, $p=10^{-2}$, $\beta=0.1$.  The $m$ values are given from Table \ref{LQCchaotTab60}.}
    \label{LQCchaot60}  
\end{figure*}

On the other hand, from Eq.(\ref{eq3.11b}) we find that
\bqn
\lb{eq3.30}
\epsilon_V &=& \frac{m^2_P}{4 \pi}\left(\frac{1-2\alpha_1\phi + 3\alpha_2\phi^2}{\phi\left(1-\alpha_1\phi + \alpha_2\phi^2\right)}\right)^2,\nb\\
\eta_V &=& \frac{m^2_P}{4\pi \phi^2(1-\alpha_1\phi + \alpha_2\phi^2)^2}\big(1-6\alpha_1\phi   +\nb\\
&& 6\left(2\alpha_2 + \alpha_1^2\right)\phi^2 
 -20\alpha_1\alpha_2\phi^3 + 15\alpha_2^2\phi^4\big). ~~~~~~~
\eqn
Then, for any given ($A_s, n_s, \alpha_1, \alpha_2$), from Eqs.(\ref{eqA.39}) and (\ref{eq3.30}) we can find $\phi_*$, with which Eq.(\ref{eqA.38}) yields the value of $H_*$. Then, from Eqs.(\ref{eq3.30}) and (\ref{eqA.40}) we can find the mass parameter $m$. Once $(H_*, \phi_*, m)$ are known, Eqs. (\ref{eqA.41})-(\ref{eqA.43}), together with the dynamical equations shall uniquely determine 
$(a_*, \dot\phi_*, k_*)$. Taking such obtained solutions $(a_*, H_*, \phi_*,\dot\phi_*, k_*, m)$ as the first order-approximation, one can perturbatively find the effects of ${\cal{F}}(k_*) \not=1, \;  \xi_{\text{nBD}}(k_*) \not= 0$, as illustrated explicitly in \cite{Yogesh:2024iip}. However, to our current purpose the first-order approximation is already sufficiently accurate, as argued above, so we shall not consider such corrections.

In addition, to calculate the probability of the inflation, it is found sufficient once the mass parameter $m$ is known, as to be shown below. Therefore, in the following, we need only to solve Eqs.(\ref{eqA.38aa})-(\ref{eqA.40aa}) for any given ($A_s, n_s, \alpha_1, \alpha_2$) to find $(H_*, \phi_*, m)$. Planck 2018 found that 
\bqn
\lb{Planck}
\left(A_s, n_s\right)_{\text{Planck}} = \left(2.0989 \times 10^{-9}, 0.9649\right),
\eqn
for the pivot mode
$k_*/a_0 = 0.05/\text{Mpc}$ \cite{Planck:2018jri}, while ACT found 
\bqn
\lb{ACT}
\left(A_s, n_s\right)_{\text{ACT}} = \left(2.1179 \times 10^{-9}, 0.9666\right),
\eqn
for the same   mode \cite{AtacamaCosmologyTelescope:2025blo}. In addition, the combination of Planck 2018 (P), ACT 2025 (ACT) and DESI DR2 (LB2) \cite{DESI:2025zpo,DESI:2025zgx} yields  
\cite{AtacamaCosmologyTelescope:2025blo}
\bqn
\left(A_s, n_s\right)_{\text{Combined}} = \left(2.1370 \times 10^{-9}, 0.9752\right),
\eqn
 which excludes various previously-favored inflationary models \cite{Kallosh:2025ijd}.

To see the effects from different observations, let us first consider the case where $\alpha_1 = 0.14$ and $\alpha_2 = 6.644\times 10^{-3}$ \cite{Kallosh:2025ijd}. Then, for each set of $(A_s, n_s)$, we find that Eqs.(\ref{eqA.38aa}) - (\ref{eqA.40aa}) have two solutions for the parameter $m$, given in Table \ref{TableI}, which corresponds to different signs of $\phi_*$ as shown in the same table. For $\phi_* > 0$ we can see that the combination of Planck + ACT + LB2  yields the smallest value of $m$, the same is for the value of $H_*$. 
For $\phi_* < 0$ a similar behavior of ($m, H_*$) is observed, although the corresponding value of $m$ is of  one-order smaller than that given in the case $\phi_* > 0$, while the values of $H_*$ are slightly larger than those given in the case $\phi_* > 0$.

With the above obtained mass $m$, we can find the total e-fold $N_{\text{inf}}$ defined by Eq.(\ref{efolds definition}). Requiring $N_{\text{inf}} \gtrsim 60$, we find that $\phi_B$ must be in the ranges
\bqn
\lb{eq3.31}
\phi_B \in \left(\phi_{\text{min}}, \phi_1\right)\cup \left(\phi_2, \phi_{\text{max}}\right),
\eqn
where $\left(\phi_{\text{min}}, \phi_1, \phi_2, \phi_{\text{max}}\right)$ are given in Table \ref{TableI} for each value of $m$.

In Fig.\ \ref{polyefolds}  
we plot the e-folds vs the initial values of $\phi_B$ for both $\dot\phi_B > 0$ and  $\dot\phi_B < 0$ with $m = 1.11\times 10^{-6}\; m_P$. From this figure we can see that inflation with $N_{\text{inf}} \gtrsim 60$ exists in both cases. In addition,  the probabilities for $\dot\phi_B > 0$ is similar to that of $\dot\phi_B < 0$.
The same is true also for other cases to be considered below. Therefore, without loss of the generality, we shall consider only the case $\dot\phi_B > 0$. Then, from Table \ref{TableI} we can see that $N_{\text{inf}} \gtrsim 60$
is achievable for 
\bq
\lb{4.10ee}
\phi_B \in 
\left(-489.64, -5.63\right) \cup \left(0.28, 503.68\right).
\eq
 Then, from Eq.(\ref{4.9}) we obtain
\bqn
\lb{4.10ff}
P_{\text{LQC}} \left(\text{not realized}\right) \lesssim
 6.52\times 10^{-3}.
\eqn
For other values of $m$, the corresponding $P_{\text{LQC}} \left(\text{not realized}\right)$ is given in Table \ref{LQCpolyTab}, from which 
we find that 
\bqn
\lb{eq3.32}
P_{\text{LQC}} \left(\text{not realized}\right) \lesssim \frac{\int_{\phi_1}^{\phi_2}{d\omega}}
{\int_{\phi_{\text{min}}}^{\phi_{\mathrm{max}}} {d\omega}}
\approx {\cal{O}}\left(10^{-3}\right),  
\eqn
which is about three-orders larger than that given by Eq.(\ref{4.10cc}) for the chaotic potential \cite{Ashtekar:2011rm}.

Choosing $\alpha_1 = 0.13, \; \alpha_2 = 5.746\times 10^{-3}$ \cite{Kallosh:2025ijd}, we find that the corresponding values of the physical quantities discussed above are not changed dramatically. In particular, we still have 
$$
P_{\text{LQC}} \left(\text{not realized}\right) \approx {\cal{O}}\left(10^{-3}\right),
$$
as shown in Table \ref{LQCpolyTab2}.

%%%%%%%%%%%%%%%%
%% Table V
%%%%%%%%%%%%%%%%
\begin{table*}[]
\begin{tabular}{|c|c|c|r|c|c|r|c|}
\hline
                          & $m$                                       & $H_*$                                     & \multicolumn{1}{c|}{$\phi_*$} & $\phi_{\text{min}}$ & $\phi_{\text{max}}$ & \multicolumn{1}{c|}{$\phi_2$} & $P_{\text{LQC}}$ \\ \hline
\multirow{2}{*}{Planck}   & $1.56\times 10^{-6}$                      & $5.55\times 10^{-6}$                      & $2.52$                        & $-0.20$             & $0.41$              & N/A                           & $1$              \\ \cline{2-8} 
                          & $6.93\times 10^{-7}$                      & $8.73\times 10^{-6}$                      & $-3.79$                       & $-0.20$             & $0.41$              & $0.27$                        & $0.97$           \\ \hline
\multirow{2}{*}{ACT}      & $1.49\times 10^{-6}$                      & $5.37\times 10^{-6}$                      & $2.57$                        & $-0.20$             & $0.41$              & N/A                           & $1$              \\ \cline{2-8} 
                          & $6.51\times 10^{-7}$                      & $8.57\times 10^{-6}$                      & $-3.90$                       & $-0.20$             & $0.41$              & $0.33$                        & $0.96$           \\ \hline
\multirow{2}{*}{Combined} & $1.11\times 10^{-6}$                      & $4.30\times 10^{-6}$                      & $2.92$                        & $-0.20$             & $0.41$              & 0.41                          & $1$              \\ \cline{2-8} 
                          & $4.31\times 10^{-7}$                      & $7.50\times 10^{-6}$                      & $-4.70$                       & $-0.20$             & $0.41$              & $0.27$                        & $0.94$           \\ \hline
\multirow{2}{*}{WMAP}     & \multicolumn{1}{l|}{$1.53\times 10^{-6}$} & \multicolumn{1}{l|}{$5.58\times 10^{-6}$} & \multicolumn{1}{c|}{$2.62$}   & $-0.20$             & $0.41$              & N/A                           & $1$              \\ \cline{2-8} 
                          & \multicolumn{1}{l|}{$6.58*10^{-7}$}       & \multicolumn{1}{l|}{$9.00\times 10^{-6}$} & \multicolumn{1}{c|}{$-4.01$}  & $-0.20$             & $0.41$              & $0.33$                        & $0.96$           \\ \hline
\end{tabular}
\caption{The physical quantities ($H_*, \phi_*, m$) in LQC for a given set of ($A_s, n_s$) from the observations, Planck, ACT, Planck + ACT + LB2 (Combined), or WMAP,  and the corresponding values of $\phi_{\text{min}}$, $\phi_{\text{max}}$, $\phi_2$, and  $P_{\text{LQC}}$ for   $\dot{\phi}_B > 0$, $\alpha_1 = 0.14, \; \alpha_2 = 6.644\times 10^{-3}$ and $U_0=10^2, p=0.011,  \beta=0.5$.}
\label{LQCpolyTab60}
\end{table*}

%%%%%%%%%%%%%%%%%%%%%
%% Fig.\ 5
%%%%%%%%%%%%%%%%%%%%%%

  \begin{figure*}[t] 
    %\centering   
    \includegraphics[width=7cm]{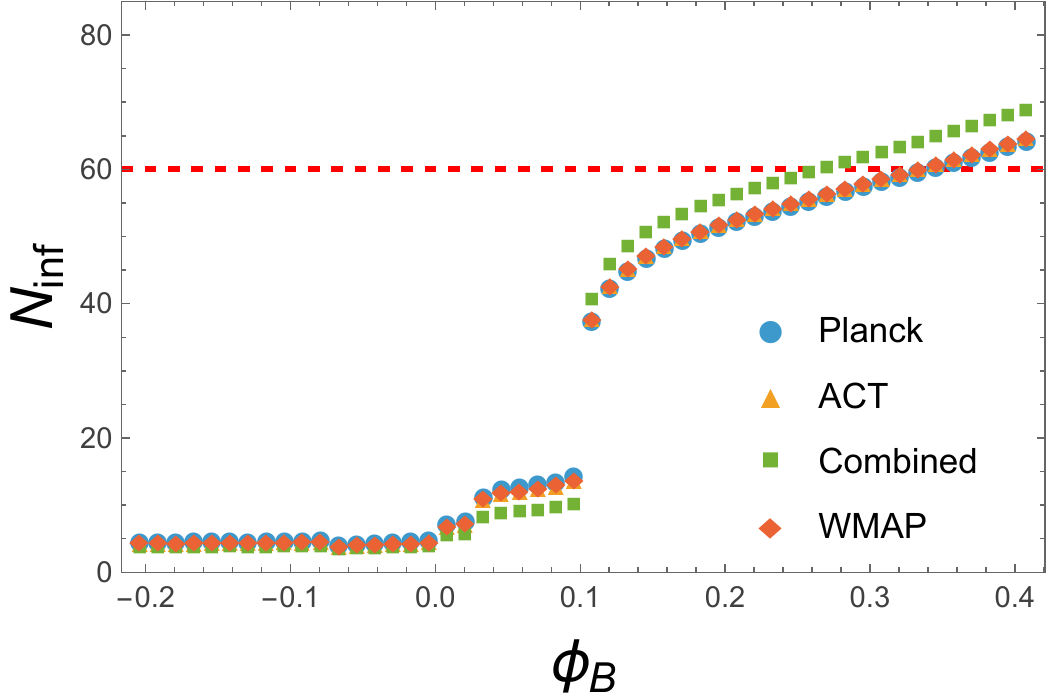} 
    \vspace{.5cm}
\caption{The plots of the total number of e-folds with a chaotic potential,  given by Eq.(\ref{eq3.11}) and an ekpyrotic potential $\alpha_1 =0.14 ,\alpha_2 = 6.644*10^{-3}$, given 
    by Eqs.(\ref{eq3.3}) in the framework of LQC for $\dot{\phi}_B >0$, where the parameters are chosen as  $U_0=100$, $p=0.011$, $\beta=0.5$.  The $m$ values are given from Table \ref{LQCpolyTab60}.}
    \label{LQCpoly60}  
\end{figure*}

%%%%%%%%%%%%%%%%
%% Table VI
%%%%%%%%%%%%%%%%

\begin{table*}[]
\centering
\resizebox{\textwidth}{!}{%
\begin{tabular}{|c|c|c|r|c|c|r|r|c|}
\hline
                          & $m$                  & $H_*$                & \multicolumn{1}{c|}{$\phi_*$} & $\phi_{\text{min}}$ & $\phi_{\text{max}}$ & \multicolumn{1}{c|}{$\phi_1$} & \multicolumn{1}{c|}{$\phi_2$} & $P_{\text{mLQC-I}}$  \\ \hline
\multirow{2}{*}{Planck}   & $1.56\times 10^{-6}$ & $5.54\times 10^{-6}$ & $2.52$                        & $-341.91$           & $355.96$            & $-5.45$                       & $0.45$                        & $9.28\times 10^{-3}$ \\ \cline{2-9} 
                          & $6.93\times 10^{-7}$ & $8.73\times 10^{-6}$ & $-3.79$                       & $-450.22$           & $464.27$            & $-5.58$                       & $0.30$                        & $7.06\times 10^{-3}$ \\ \hline
\multirow{2}{*}{ACT}      & $1.49\times 10^{-6}$ & $5.37\times 10^{-6}$ & $2.58$                        & $-347.10$           & $361.15$            & $-5.45$                       & $0.43$                        & $9.11\times 10^{-3}$ \\ \cline{2-9} 
                          & $6.51\times 10^{-7}$ & $8.57\times 10^{-6}$ & $-3.90$                       & $-459.89$           & $473.94$            & $-5.60$                       & $0.30$                        & $6.94\times 10^{-3}$ \\ \hline
\multirow{2}{*}{Combined} & $1.11\times 10^{-6}$ & $4.30\times 10^{-6}$ & $2.92$                        & $-528.40$           & $542.45$            & $-5.50$                       & $0.40$                        & $8.29\times 10^{-3}$ \\ \cline{2-9} 
                          & $4.31\times 10^{-7}$ & $7.51\times 10^{-6}$ & $-4.70$                       & $-692.69$           & $707.77$            & $-5.65$                       & $0.23$                        & $6.03\times 10^{-3}$ \\ \hline
\end{tabular}%
}
\caption{The physical quantities ($H_*, \phi_*, m$) for a given set of ($A_s, n_s$) from the observations, Planck, ACT or Planck + ACT + LB2 (Combined),  and the corresponding values of $\phi_{\text{min}}$, $\phi_{\text{max}}$, $\phi_1$, $\phi_2$, and  $P_{\text{mLQC-I}}$ for   $\dot{\phi}_B > 0$, and $\alpha_1 = 0.14$ and $\alpha_2 = 6.644\times 10^{-3}$ \cite{Kallosh:2025ijd} in mLQC-I.}
\label{mLQCpolyTab}
\end{table*}

%%%%%%%%%%%%%%%%%%%%%
%% Table VII
%%%%%%%%%%%%%%%%%%%%%%
\begin{table*}[]
\centering
\resizebox{\textwidth}{!}{%
\begin{tabular}{|c|c|c|r|c|c|r|r|c|}
\hline
                          & $m$                  & $H_*$                & \multicolumn{1}{c|}{$\phi_*$} & $\phi_{\text{min}}$ & $\phi_{\text{max}}$ & \multicolumn{1}{c|}{$\phi_1$} & \multicolumn{1}{c|}{$\phi_2$} & $P_{\text{mLQC-I}}$  \\ \hline
\multirow{2}{*}{Planck}   & $1.55\times 10^{-6}$ & $5.72\times 10^{-6}$ & $2.55$                        & $-359.09$                                  & $374.17$            & $-5.43$                       & $0.48$                        & $8.83\times 10^{-3}$ \\ \cline{2-9} 
                          & $7.28\times 10^{-7}$ & $8.68\times 10^{-6}$ & $-3.72$                       & $-464.56$                                  & $479.64$            & $-5.55$                       & $0.35$                        & $6.86\times 10^{-3}$ \\ \hline
\multirow{2}{*}{ACT}      & $1.49\times 10^{-6}$ & $5.55\times 10^{-6}$ & $2.60$                        & $-364.47$                                  & $379.55$            & $-5.43$                       & $0.48$                        & $8.71\times 10^{-3}$ \\ \cline{2-9} 
                          & $6.85\times 10^{-7}$ & $8.52\times 10^{-6}$ & $-3.83$                       & $-474.29$                                  & $489.38$            & $-5.55$                       & $0.35$                        & $6.72\times 10^{-3}$ \\ \hline
\multirow{2}{*}{Combined} & $1.11\times 10^{-6}$ & $4.49\times 10^{-6}$ & $2.95$                        & $-402.06$                                  & $417.14$            & $-5.48$                       & $0.43$                        & $7.91\times 10^{-3}$ \\ \cline{2-9} 
                          & $4.59\times 10^{-7}$ & $7.46\times 10^{-6}$ & $-4.62$                       & $-543.17$                                  & $558.25$            & $-5.63$                       & $0.28$                        & $5.88\times 10^{-3}$ \\ \hline
\end{tabular}%
}
\caption{The physical quantities ($H_*, \phi_*, m$) for a given set of ($A_s, n_s$) from the observations, Planck, ACT or Planck + ACT + LB2 (Combined),  and the corresponding values of $\phi_{\text{min}}$, $\phi_{\text{max}}$, $\phi_1$, $\phi_2$, and  $P_{\text{mLQC-I}}$ for $\dot{\phi}_B > 0$, and $\alpha_1 = 0.13$ and $\alpha_2 = 5.746\times 10^{-3}$ \cite{Kallosh:2025ijd} in mLQC-I.}
\label{mLQCpolyTab2}
\end{table*}

%%%%%%%%%%%%%%%%%%%%%
%% Fig.\ 6
%%%%%%%%%%%%%%%%%%%%%%

\begin{figure*}[t!] 
    \centering
   \includegraphics[width=7cm]{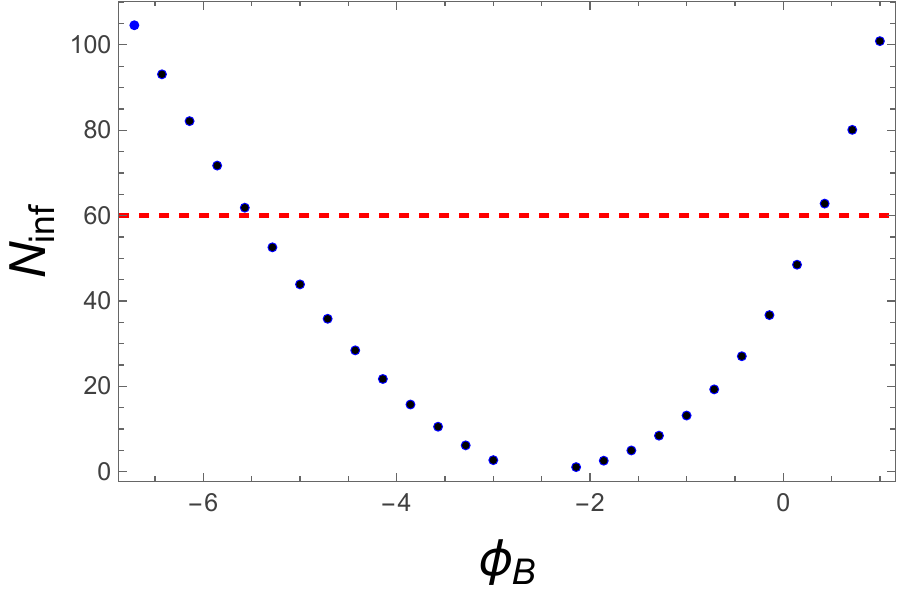} 
    \includegraphics[width=7cm]{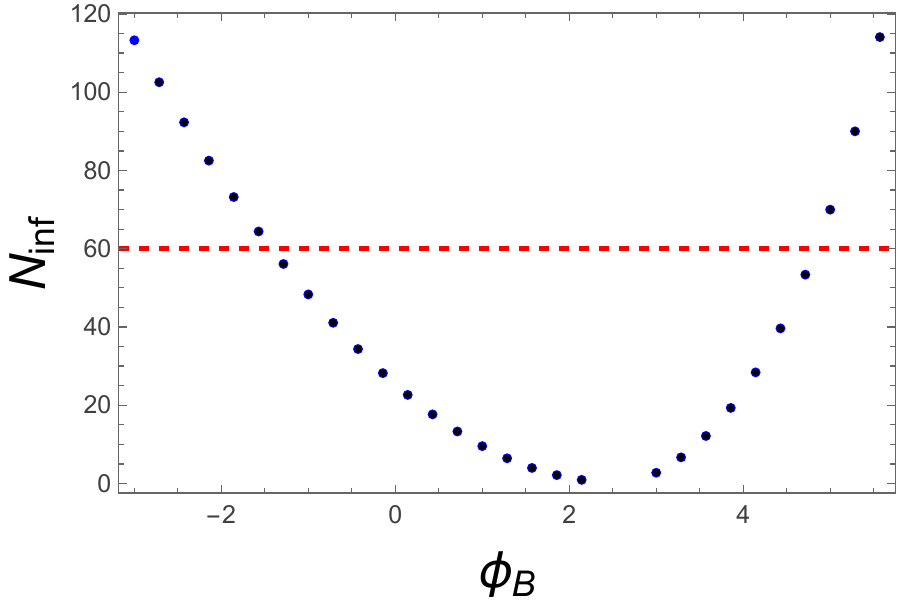} \\  
    \vspace{.2cm}
   \mbox{\hspace{1.cm}}  (a) \mbox{\hspace{6.8cm}} (b)\\
    \caption{The plots of the total number of e-folds  during the inflationary phase in mLQC-I, when the  polynomial chaotic potential given by Eq.(\ref{eq3.11}) is  present with the choice $\alpha_1 = 0.14, \; \alpha_2 = 6.644 \times 10^{-3}$ and $m = 1.11 \times 10^{-6}\; m_\text{pl}$. Plots (a) and (b) are respectively for $\dot\phi_B >0$ and   $\dot\phi_B < 0$.}
    \label{fig6} %mlqcpolyefolds}
\end{figure*}

%%%%%%%%%%%%%%%%
%% Table VIII
%%%%%%%%%%%%%%%%
\begin{table*}[]
\begin{tabular}{|c|c|c|r|c|c|r|c|}
\hline
                          & $m$                                       & $H_*$                                     & \multicolumn{1}{c|}{$\phi_*$} & $\phi_{\text{min}}$ & $\phi_{\text{max}}$ & \multicolumn{1}{c|}{$\phi_2$} & $P_{\text{mLQC-I}}$ \\ \hline
\multirow{2}{*}{Planck}   & $1.56\times 10^{-6}$                      & $5.55\times 10^{-6}$                      & $2.52$                        & $-0.23$             & $0.45$              & N/A                           & $1$              \\ \cline{2-8} 
                          & $6.93\times 10^{-7}$                      & $8.73\times 10^{-6}$                      & $-3.79$                       & $-0.23$             & $0.45$              & N/A                        & $1$           \\ \hline
\multirow{2}{*}{ACT}      & $1.49\times 10^{-6}$                      & $5.37\times 10^{-6}$                      & $2.57$                        & $-0.23$             & $0.45$              & N/A                           & $1$              \\ \cline{2-8} 
                          & $6.51\times 10^{-7}$                      & $8.57\times 10^{-6}$                      & $-3.90$                       & $-0.23$             & $0.45$              & $0.45$                        & $1$           \\ \hline
\multirow{2}{*}{Combined} & $1.11\times 10^{-6}$                      & $4.30\times 10^{-6}$                      & $2.92$                        & $-0.23$             & $0.45$              & N/A                          & $1$              \\ \cline{2-8} 
                          & $4.31\times 10^{-7}$                      & $7.50\times 10^{-6}$                      & $-4.70$                       & $-0.23$             & $0.45$              & $0.40$                        & $0.99$           \\ \hline
\multirow{2}{*}{WMAP}     & \multicolumn{1}{l|}{$1.53\times 10^{-6}$} & \multicolumn{1}{l|}{$5.58\times 10^{-6}$} & \multicolumn{1}{c|}{$2.62$}   & $-0.23$             & $0.45$              & N/A                           & $1$              \\ \cline{2-8} 
                          & \multicolumn{1}{l|}{$6.58*10^{-7}$}       & \multicolumn{1}{l|}{$9.00\times 10^{-6}$} & \multicolumn{1}{c|}{$-4.01$}  & $-0.23$             & $0.45$              & $0.45$                        & $1$           \\ \hline
\end{tabular}
\caption{The physical quantities ($H_*, \phi_*, m$) in  mLQC-I for a given set of ($A_s, n_s$) from the observations, Planck, ACT, Planck + ACT + LB2 (Combined), or WMAP,  and the corresponding values of $\phi_{\text{min}}$, $\phi_{\text{max}}$, $\phi_2$, and  $P_{\text{LQC}}$ for   $\dot{\phi}_B > 0$, $\alpha_1 = 0.14, \; \alpha_2 = 6.644\times 10^{-3}$ and $U_0=10^2, p=0.011,  \beta=0.5$.}
\label{mLQCpolyTab60}
\end{table*}

 %%%%%%%%%%%%%%%%%%%%%
%% Fig.\ 7
%%%%%%%%%%%%%%%%%%%%%%

\begin{figure*}[t] 
    \centering    
    \includegraphics[width=7cm]{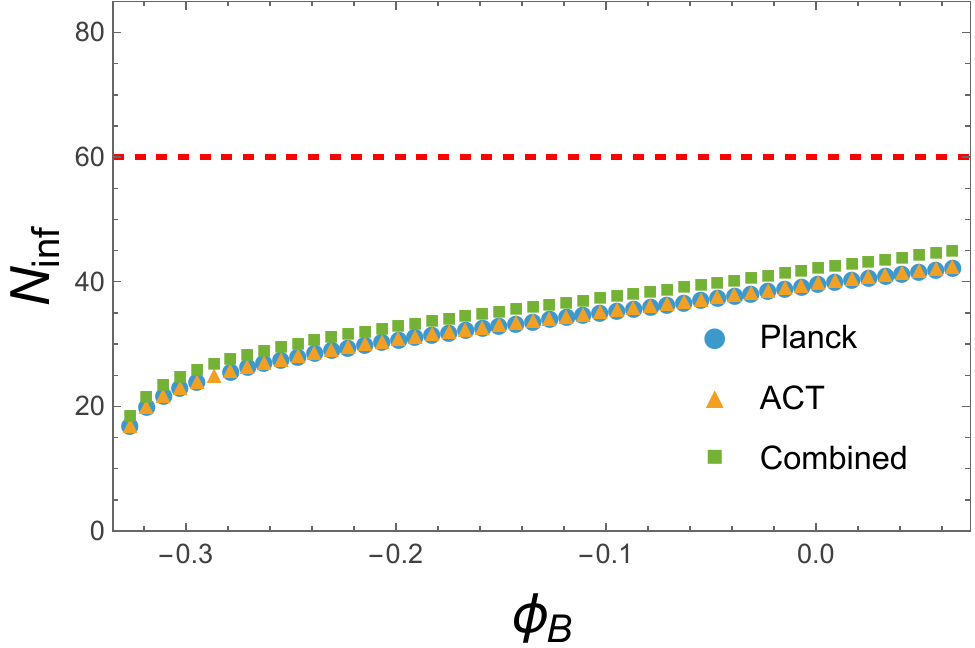} 
     \includegraphics[width=7cm]{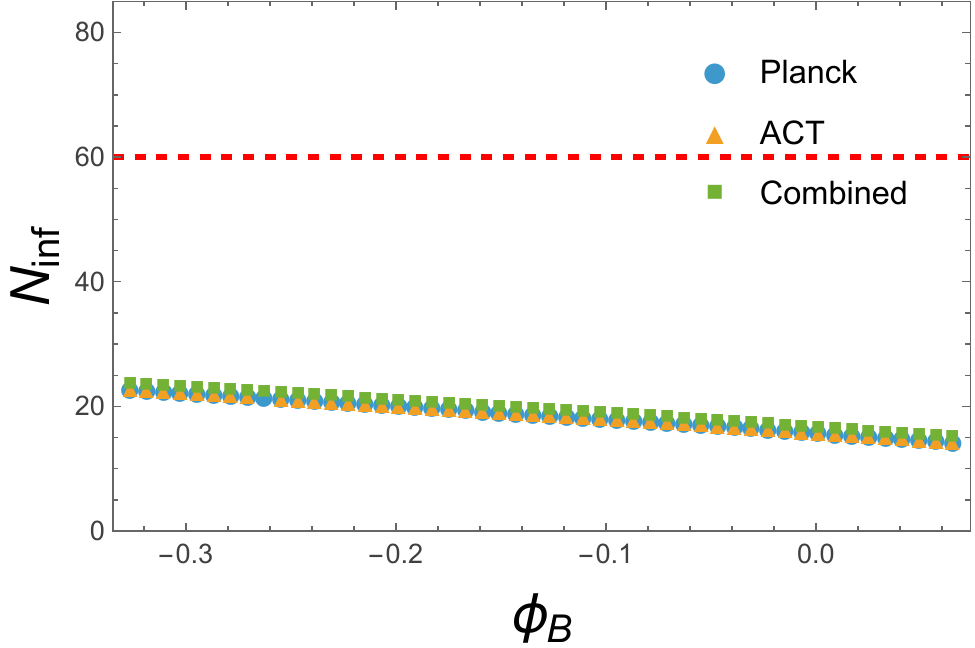}\\  
    %\vspace{.2cm}
   \mbox{\hspace{1.cm}}  (a) \mbox{\hspace{6.8cm}} (b)\\
    \vspace{.5cm}
    \caption{The plots of the total number of e-folds with a polynomial potential, $\alpha_1 =0.14 ,\alpha_2 = 6.644*10^{-3}$, given by Eq.(\ref{eq3.11}) and an ekpyrotic potential given 
    by Eqs.(\ref{eq3.3}) in the framework of mLQC-I, where the parameters are chosen as  $U_0=0.0366$, $p=0.1$, $\beta=5$.  The plots (a) and (b) are  for $\dot\phi_B>0$ and  $\dot\phi_B< 0$, respectively. Three cases are plotted on each graph, corresponding to the three different  masses given in Table \ref{mLQCpolyTab60}. Note that in the figure the lines for Planck and ACT are practically indistinguishable.}
    \label{mLQCpolySingh}  
\end{figure*}

%%%%%%%%%%%%%%%%%%%%%
%% Fig.\ 8
%%%%%%%%%%%%%%%%%%%%%%

\begin{figure*}[t] 
    %\centering   
    \includegraphics[width=7cm]{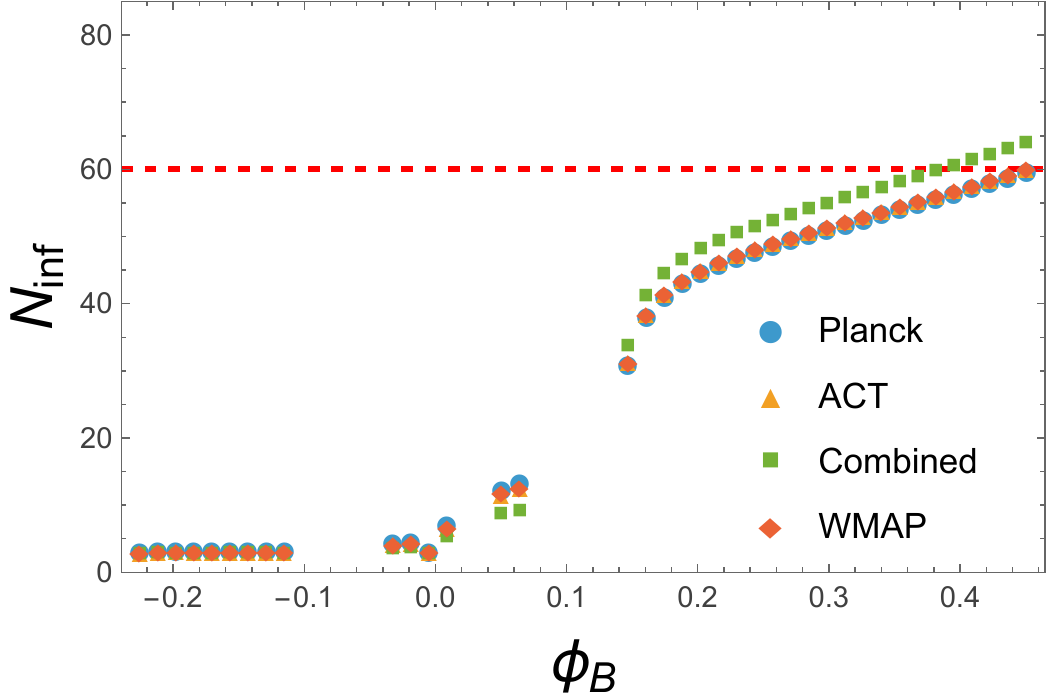} 
    \vspace{.5cm}
    \caption{The plots of the total number of e-folds with a chaotic potential, $\alpha_1 =0.14 ,\alpha_2 = 6.644*10^{-3}$,  given by Eq.(\ref{eq3.11}) and an ekpyrotic potential given 
    by Eqs.(\ref{eq3.3}) in the framework of mLQC-I for $\dot{\phi}_B >0$, where the parameters are chosen as  $U_0=100$, $p=0.011$, $\beta=0.5$.  and the $m$ values are given from Table \ref{mLQCpolyTab60}.}
    \label{mLQCpoly60}  
\end{figure*}

\subsection{Effects of the Ekpyrotic Mechanism}

Following what we did in the last subsection, 
let us start with the choice of the parameters    $U_0=0.0366$, $p=0.1$, $\beta=5$    for the ekpyrotic potential considered in \cite{McNamara:2022dmf}, while  setting  $\alpha_1 = \alpha_2 = 0$. Then, solving Eqs.(\ref{eqA.38aa})-(\ref{eqA.40aa}) we find $\left(H_*, \phi_*, m\right)$ for a given set of $\left(A_s, n_s\right)$ from observations. In Table \ref{LQCchaotTabSingh} we list the corresponding values of $\left(H_*, \phi_*, m\right)$ for the Planck, ACT and Planck + ACT + LB2, respectively. With such obtained mass parameter $m$, we solve Eq.(\ref{eq3.6})  with $W_{\text{Bmin}} = 1.001$ and find that the corresponding $\phi_{\text{Bmin}}$ and $\phi_{\text{Bmax}}$, where $\phi_{\text{Bmax}}$ is given by Eq.(\ref{eq3.6a}). It is interesting to note that   $\phi_{\text{Bmin}}$ and $\phi_{\text{Bmax}}$ are the same  within our error bars in the three cases. Then, the range of $\phi_B$ is   
\bqn
\lb{eq3.9aa}
\phi_B \in \left(-0.26, 0.05\right),
\eqn
in all three cases.

Comparing Table \ref{LQCchaotTabSingh} with Tables \ref{LQCpolyTab} and \ref{LQCpolyTab2}, one finds the dramatic difference between the range of $\phi \in \left(\phi_{\text{min}},\phi_{\text{max}}\right)$, given in the case without the ekpyrotic potential, and the range of $\phi \in  \left(\phi_{\text{Bmin}},\phi_{\text{Bmax}}\right)$
when the ekpyrotic potential is present. In particular,  the range of $\phi \in \left(\phi_{\text{min}},\phi_{\text{max}}\right)$ is determined by Eq.(\ref{phiMN}), while 
the range of $\phi \in  \left(\phi_{\text{Bmin}},\phi_{\text{Bmax}}\right)$ is determined by Eqs.(\ref{eq3.6}) and (\ref{eq3.6a}) for a given $w_{B} > 1$.

With the mass parameter given in Table \ref{LQCchaotTabSingh},  in Fig.\ \ref{LQCchaotSingh} we plot $N_{\text{inf}}$ for $\phi_{B} \in\left(\phi_{\text{Bmin}}, \phi_{\text{Bmax}} \right)$. From this figure it can be seen clearly that the e-fold $N_{\text{inf}}$ during the inflationary phase is now always $N_{\text{inf}} < 40$,  due to the effects of the  ekpyrotic mechanism. This is true not only for the mass parameter obtained from Planck 2018, but also true for ACT 2025 and the combined one. It should be noted that in Fig.\ \ref{LQCchaotSingh} the lines for Planck and ACT are practically  indistinguishable.

However, when choosing different values of the ekpyrotic potential parameters, we can get e-folds larger than $60$. For example, choosing   $U_0=10^{-3}, \; p=10^{-2},\; \beta=0.1$, 
in Table \ref{LQCchaotTab60} we show the relevant physical quantities. From this table  we can see that the parameters $m$, $\phi_{60}$ and $P_{\text{LQC}}(\text{not realized})$ sensitively depend on the observations, while the ones $w_{\text{Bmin}}$, $w_{\text{Bmax}}$,  $\phi_{\text{Bmin}}$, and $\phi_{\text{Bmax}}$ are indistinguishable.  In this table,  we also included the WMAP data \cite{2011ApJS..192...18K}, $\left(A_s, n_s\right) =  \left(2.43\times10^{-9}, 0.968\right)$, adopted in   \cite{Ashtekar:2011rm}.

In Fig.\ \ref{LQCchaot60} we plot $N_{\text{inf}}$ vs $\phi_B$, from which  we can see that WMAP and Planck give almost the same e-fold for any given $\phi_B$. In particular, the corresponding $\phi_{60}$'s are respectively $0.099$ and $0.105$. On the other hand,  ACT yields $\phi_{60} \simeq 0.170$, and the corresponding $P_{\text{LQC}}(\text{not realized}) \simeq 57.2\%$, while for the combined one, $N_{\text{inf}}$ is always less than $60$, so such chosen parameters are already ruled out by the combined data, which is somehow unexpected. In any case, comparing it with the cases without   the  ekpyrotic mechanism, we can see that the probabilities for inflation to occur now is much smaller
\bqn
\lb{eq3.43}
P_{\text{LQC}}\left(\text{not realized}\right) \approx {\cal{O}}(1).
\eqn
Therefore, even now initial conditions with $N_{\text{inf}} \gtrsim 60$ exists, for such inflation to occur, fine-tuning is still required.

When $\alpha_1\alpha_2 \not=0$, we find that for the choice of the parameters   $U_0=0.0366$, $p=0.1$, $\beta=5$   \cite{McNamara:2022dmf} and $\alpha_1 = 0.14,\; \alpha_2 = 6.644\times 10^{-3}$, inflation with $N_{\text{inf}} \gtrsim 60$ does not exist, similar to the previous cases. 
But, if we choose $U_0 = 10^2,\; p = 0.011, \; \beta = 0.5$ and $\alpha_1 = 0.14,\; \alpha_2 = 6.644\times 10^{-3}$, inflation with $N_{\text{inf}} \gtrsim 60$ does exist. In particular, we find that, for a given set of the observational values $(A_s, n_s)$, there exist two masses, as given in Table \ref{LQCpolyTab60}. Then, the corresponding e-fold for the first value of $m$ given in Table \ref{LQCpolyTab60} in each case is always less than $60$, while for the second   value of $m$ inflation with  $N_{\text{inf}} \gtrsim 60$ exists in each case, as shown in Fig.\ \ref{LQCpoly60}. But, again the
probability for inflation to occur is still given by Eq.(\ref{eq3.43}), as can be seen from Table \ref{LQCpolyTab60}.  
Therefore,   for desired  inflation to occur, fine-tuning is also required.

We have also considered many other choices of ($U_0$, $p$ $\beta$) and not been able to find any choice that gives a desired inflation with $N_{\text{inf}} \gtrsim 60$ and $P_{\text{LQC}}\left(\text{not realized}\right) \ll 1$.

In the next section, we consider the effects of the mechanism on inflation in mLQC-I and obtain similar conclusions.

\section{Effects of the Ekpyrotic Mechanism on Inflation in mLQC-I}
\renewcommand{\theequation}{4.\arabic{equation}}
\setcounter{equation}{0}
\lb{SecIV}

 Similar to the last section,  let us consider the cases  with and without the ekpyrotic mechanism separately.

 \subsection{Probability of Inflation  Without  Ekpyrotic Mechanism}

Without the  ekpyrotic mechanism, 
the studies of the probability of inflation were further generalized to modified LQCs \cite{Li:2019ipm}, in which it was found that the critical energy density now is given by $\rho_c^{\text{I}}$, given by Eq.(\ref{eq_rho}). Therefore, to find $\phi_{\text{min}}$ and $\phi_{\text{max}}$ from Eq.(\ref{phiMN}), we need first to replace $\rho_c$ by $\rho_c^{\text{I}}$.
 For the chaotic potential, $\alpha_1 = \alpha_2 = 0$,  it was found that to have $N_{\text{inf}} \gtrsim 60$ the initial values of $\phi_B$ must be taken from the intervals 
\bqn
\lb{eq3.36}
\phi_B \in \left(-\phi_{\text{max}}, -5.518\; m_P\right) \cup \left(0.917\; m_P, \phi_{\text{max}}\right),
\eqn
where $\phi_{\text{max}} = 8 3.49\times 10^5\; m_P$. Then, the probability that the desired slow-roll inflation does not happen is given by \cite{Li:2019ipm}
\bqn
\lb{eq3.37}
P_{\text{mLQC-I}} \left(\text{not realized}\right) &\lesssim& \frac{\int_{-5.158 m_P}^{0.917 m_P}{d\omega}}
{\int_{-\phi_{\text{max}}}^{\phi_{\mathrm{max}}} {d\omega}}
\lesssim
1.12\times 10^{-5},\nb\\
\eqn
which is comparable to that given by Eq.(\ref{4.10ff}).

However, when $\alpha_1\alpha_2 \not= 0$, similar to those studied in LQC, now for each given set of $(A_s, n_s)$, Eqs.(\ref{eqA.38aa})-(\ref{eqA.40aa}) have two real solutions for the parameter $m$, as shown in 
 Tables \ref{mLQCpolyTab} and \ref{mLQCpolyTab2}, respectively, for $\left(\alpha_1, \alpha_2\right) = \left(0.14, 6.644\times 10^{-3}\right)$ and $\left(\alpha_1, \alpha_2\right) = \left(0.13,5.746\times 10^{-3}\right)$.
 In Fig.\ \ref{fig6}, we show the e-folds of inflation for the choice of $\left(\alpha_1, \alpha_2\right) = \left(0.14, 6.644\times 10^{-3}\right)$, from which we can read off $\phi_1$ and $\phi_2$. In   Tables \ref{mLQCpolyTab} and \ref{mLQCpolyTab2}, we provide the values of $\phi_1$ and $\phi_2$ for each given $m$. 
 From these tables we find %that now we also have   
 \bq
 \lb{eq3.37}
 P_{\text{LQC}} \left(\text{not realized}\right) \lesssim  
{\cal{O}}\left(10^{-3}\right),
\eq
which is quite similar to that studied in the last subsection in LQC.

\subsection{Probability of Inflation  With  Ekpyrotic Mechanism}

Let us again first consider the parameters 
$U_0=0.0366$, $p=0.1$, $\beta=5$, with $m = 1.26 \times 10^{-6}\; m_{\text{pl}}$ and  $\alpha_1 = 0.14, \; \alpha_2 = 6.644 \times 10^{-3}$ for the chaotic potential. Then, from Eqs.(\ref{eqA.38aa})-(\ref{eqA.40aa}) we find that the mass parameter $m$ is given by Table \ref{mLQCpolyTab60} for every given set of $(A_s, n_s)$, due to the fact that these equations depend only on the potential not on the model. Then, in Fig.\ \ref{mLQCpolySingh} we plot out the corresponding e-folds, respectively, for Planck, ACT and the combined one. From this figure we can see that the ekpyrotic mechanism affects the e-folds of inflation dramatically. In particular, now $N_{\text{inf}}$ is always less than $40$, sharply in contrast to the case without the ekpyrotic mechanism, shown   in Fig.\ \ref{fig6}.

%%%%%%%%%%%%%%%%%
%% Fig.\ 7

%%%%%%%%%%%%%%%%%%%%%%%%%%%%%%%%%%%%%%%%%%%%%%%%%%%%%%%%%%%%%%%%

However, when adjusting the parameters of the ekpyrotic potential, we find that we can get  e-folds larger than 60. In particular, choosing $U_0=10^2, p=0.011$, and $\beta=0.5$, we first obtain the mass parameter and present it in Table \ref{mLQCpolyTab60} for different observational values of ($A_s, n_s$). The e-folds corresponding to these masses are shown in Fig.\ \ref{mLQCpoly60}. As can be seen from this figure, the Combined case easily passes 60 e-folds, while the ACT and WMAP cases barely do, with Planck still falling just short.

In this table we also show %find %that now we also have   
 \bq
 \lb{eq4.4}
 P_{\text{mLQC-I}} \left(\text{not realized}\right) \lesssim  
{\cal{O}}\left(1\right),
\eq
which is quite similar to that studied in the last subsection in LQC.

\section{Conclusions and Remarks}
\renewcommand{\theequation}{4.\arabic{equation}}
\setcounter{equation}{0}
\lb{SecIV}

Inflation is generic in both LQC \cite{Ashtekar:2011rm} and mLQCs \cite{Li:2019ipm}. However, it is not clear how shear will affect the above conclusion, as it is well-known that shear always collapses effectively as $1/a^6$ \cite{Ryan:1975jw}, which can dominate the evolution of the universe near the quantum bounce over all other matter fields, a possible exception is  the stiff fluid (or massless scalar field) \cite{McNamara:2022dmf}.
Even in the latter, it is not clear how to ensure that the stiff fluid always dominates the evolution, as both of them grow as $a^{-6}$ towards the bounce. If the shear dominates the contraction, the universe will become highly anisotropic after the bounce, whereby the assumption of the cosmological principle will be violated. A common mechanism to solve the shear problem either in classical or quantum bouncing cosmological models \cite{Lehners:2008vx,Battefeld:2014uga,Brandenberger:2016vhg,Ashtekar:2011ni,Li:2023dwy,Agullo:2023rqq} is 
to introduce an ekpyrotic type of potentials \cite{Khoury:2001wf,Lehners:2008vx},   which becomes negative near the bounce, so the effective equation of state (EoS) of the scalar field will be greater than one, whereby dominates the shear and other matter fields in the bounce region. As a result, a homogeneous and isotropic universe can be produced after the bounce.

In this paper, we have studied the effects of the ekpyrotic mechanism on the inflationary phase in LQC and mLQC-I, in which the inflation is generic \cite{Ashtekar:2011rm,Li:2019ipm} without considering the  ekpyrotic mechanism. To study such effect, we have assumed that the potential of an inflationary field 
$\phi$ consists of two parts, as those given by Eq.(\ref{eq1.1}).
 To be specific, we have taken them as given respectively by Eqs.(\ref{eq3.3}) and (\ref{eq3.11}). 
 By numerically solving the corresponding dynamical equations in the framework of both LQC and mLQC-I, we have found that the effects are dramatic. In particular, initial conditions that led to inflation with sufficient e-folds now become impossible after the ekpyrotic mechanism is taken into account although by properly choosing the free parameters involved in the models and different initial conditions, we have shown that viable inflationary models  still exist.

 However, our calculations of the probability for inflation to occur  suggest that fine-tuning may be required. Nevertheless, because the results are numerical, our conclusions are not definitive, and a more systematic analysis will be necessary to fully assess the generality of the observed behavior.

\begin{acknowledgments}

  C.B. and B.P. are supported by the Baylor Physics graduate program, and  A.W. is partially supported by the US NSF grant: PHY-2308845.

\end{acknowledgments}

\bibliographystyle{apsrev4-1}
\bibliography{Inflation_Ekpyrotic_Potentials_v7}

\end{document}